\documentclass[conference]{IEEEtran}
\IEEEoverridecommandlockouts
\usepackage{cite}
\usepackage{amsmath,amssymb,amsfonts}
\usepackage{algorithmic}
\usepackage{graphicx}
\usepackage{textcomp}
\usepackage{xcolor}
\usepackage{subcaption}
\usepackage{listings}
\usepackage{enumitem}
\usepackage[ruled]{algorithm2e}
\usepackage{caption}
\DeclareCaptionFont{8pt}{\fontsize{8pt}{6pt}\selectfont}
\def\BibTeX{{\rm B\kern-.05em{\sc i\kern-.025em b}\kern-.08em
    T\kern-.1667em\lower.7ex\hbox{E}\kern-.125emX}}
\begin{document}

\makeatletter % changes the catcode of @ to 11
\newcommand{\linebreakand}{%
  \end{@IEEEauthorhalign}
  \hfill\mbox{}\par
  \mbox{}\hfill\begin{@IEEEauthorhalign}
}
\makeatother % changes the catcode of @ back to 12

\title{DAOS as HPC Storage: a View From Numerical Weather Prediction}

\author{
\IEEEauthorblockN{Nicolau Manubens}
\IEEEauthorblockA{\textit{European Centre for Medium-Range} \\
\textit{Weather Forecasts, Germany} \\
nicolau.manubens@ecmwf.int}
\and
\IEEEauthorblockN{Tiago Quintino}
\IEEEauthorblockA{\textit{European Centre for Medium-Range} \\
\textit{Weather Forecasts, United Kingdom} \\
tiago.quintino@ecmwf.int}
\and
\IEEEauthorblockN{Simon D. Smart}
\IEEEauthorblockA{\textit{European Centre for Medium-Range} \\
\textit{Weather Forecasts, United Kingdom} \\
simon.smart@ecmwf.int}
\linebreakand
\IEEEauthorblockN{Emanuele Danovaro}
\IEEEauthorblockA{\textit{European Centre for Medium-Range} \\
\textit{Weather Forecasts, United Kingdom} \\
emanuele.danovaro@ecmwf.int}
\and
\IEEEauthorblockN{Adrian Jackson}
\IEEEauthorblockA{\textit{The University of Edinburgh, United Kingdom} \\
a.jackson@epcc.ed.ac.uk}
}

\maketitle

\begin{abstract}
Object storage solutions potentially address long-standing performance issues with POSIX file systems for certain I/O workloads, and new storage technologies offer promising performance characteristics for data-intensive use cases. %which can leverage non-volatile storage devices.
%A range of application areas, from machine learning to Numerical Weather Prediction (NWP) simulations, are sensitive to I/O performance and scaling, with I/O volumes and requirements increasing significantly over time.

In this work, we present a preliminary assessment of Intel’s Distributed Asynchronous Object Store (DAOS), an emerging high-performance object store, in conjunction with non-volatile storage and evaluate its potential use for HPC storage. We demonstrate DAOS can provide the required performance, with bandwidth scaling linearly with additional DAOS server nodes in most cases, although choices in configuration and application design can impact achievable bandwidth. We describe a new I/O benchmark and associated metrics that address object storage performance from application-derived workloads.% that can be utilised to explore real-world performance for this new class of storage systems. 
\end{abstract}


\section{Introduction}

%\subsection{Motivation}
Storage performance is increasingly important for large-scale applications. This is coupled with the rise in application categories where ingestion or production of large amounts of data is common, machine learning being an obvious example. As systems scale to ever large sizes, and a subset of applications require ever large amounts of I/O bandwidth or metadata performance, there is a significant challenge to improve the capabilities of the data storage technologies being utilised for I/O operations.
%Novel high-performance object storage solutions and non-volatile memory (NVM) are of two areas of technology development that have the potential to enable I/O performance at scale.

Object stores, on one hand, are a candidate to address long-standing scalability issues in POSIX file systems for large-scale parallel I/O applications where an intensive use of metadata operations is required, for example if operating with small-sized data files, of the order of KiBs up to a few MiBs. These scalability issues stem from constraints imposed by POSIX file system semantics, namely metadata prescriptiveness (per-file creation date, last-access date, permissions, etc.), and excessive consistency assurance\cite{nextplatform_lockwood}. Object stores provide alternative, less restrictive, semantics to POSIX file systems and, since they have been designed from scratch, unbound from existing file system implementations or standards, many of them have been able to follow radically different approaches.

Non-volatile memory (NVM), on the other hand, is a form of memory that offers a promising trade-off in data-intensive use-cases, with the potential to be used as NVRAM (also known as storage class memory, SCM) where performance and functionality is similar (albeit slower) to DRAM, and also heavily used in high-performance storage devices, such as NVMe drives.

The Distributed Asynchronous Object Store (DAOS)\cite{daos-scfa2022} is an emerging high-performance object store which features full user-space operation; use of a RAFT-based consensus algorithm for efficient, distributed, transactional indexing; and efficient byte-addressable access to NVM devices. It is being developed by Intel, and it has gained traction over the past years after consistently scoring leading positions in the I/O 500 benchmark\cite{io500}.

In this paper we investigate exploiting such technologies to provide I/O functionality for applications, considering generic benchmarking on one side to evaluate what can be achieved, but also with a focus on large-scale Numerical Weather Prediction (NWP) on the other side, to evaluate whether these technologies could be a good replacement for the current HPC storage system at the European Centre for Medium-Range Weather Forecasts (ECMWF), and potentially for other applications, users, and centres.

We present a preliminary suitability and performance assessment of DAOS in conjunction with non-volatile storage hardware, conducted as a proof-of-concept, decoupled from the complex software stack currently used at the ECMWF, but specially designed to mimic the I/O characteristics. A research HPC system that is composed of nodes with 3 TiB of Intel's Optane Data Centre Persistent Memory Modules (DCPMMs) has been employed to run benchmarks and measure performance, with a number of software approaches investigated, including using the DAOS C API directly.

\section{Related Material}

There has been recent research in object store developments and deployments for high-performance I/O, including CEPH \cite{ceph-pdp19} and CORTX Motr \cite{mero-cf18}. However, these have so far seen less adoption for very intensive data creation or processing workloads in large-scale HPCs. DAOS, which is part of what we evaluate in this paper, is one of the first production-ready object storage technologies targeting HPC, with excellent results in recent IO-500 benchmarks\cite{io500-sc2022}.

The meteorological community has investigated using custom object stores for I/O operations\cite{fdb-pasc17}, including evaluating such object store approaches using NVM technologies\cite{fdb-pasc19}. These investigations have demonstrated the performance and functionality benefits NVM and object storage can provide for NWP simulation I/O and data processing. When ported directly to SCM, these object stores have achieved up to 70 GB/s write bandwidth when used with SCM on a reduced set of I/O server nodes, providing confidence that SCM can be used to accelerate object storage technologies. However, direct porting relies on custom management of objects in NVM, which entails a high development and maintenance overhead and potentially limited usability for those adopting such approaches. Domain-agnostic NVM-capable object stores like DAOS simplify the use of NVM in production environments and enable implementation of high-performance user-facing tools, including domain-specific object stores, file system interfaces, programming language interfaces, and administrative tools.

Others have explored creating ad-hoc file systems on compute nodes with SCM or NVMe devices to exploit the local storage whilst providing the familiar file system interface applications are generally designed to use\cite{gekko-jcst2020} \cite{chfs-hpcasia2022} \cite{io500-sc2022}. Further work has been undertaken to integrate such file systems with batch schedulers and other tools\cite{norns-ieeecluster2019}. These all demonstrate the potential benefits to exploiting in-node storage for HPC applications, and the benefits to localising I/O compared to traditional parallel file systems. However, they do not provide object-centric interfaces that have been demonstrated to be beneficial for both functionality and performance for application domains such as NWP. They are still restricted to providing I/O through block-based file system approaches, which place limitations on achievable I/O performance for small I/O operations or for searching and extracting data within a large data set.

Very recent work by the authors has investigated the generic I/O performance of DAOS and Lustre on non-volatile memory systems, and has demonstrated that Lustre and DAOS can both achieve excellent I/O performance using high-performance storage hardware. However, Lustre struggles to handle the large volumes of metadata operations required to achieve high I/O bandwidth under intensive I/O workloads with small data transfer sizes. This highlights some of the potential benefits that object storage software can provide as opposed to traditional HPC file systems when looking at object-like I/O operations\cite{daos-pdsw22}.

The work presented in this paper builds on previous research in the areas of exploiting object storage and NVM technologies for HPC I/O. However, we are looking at the requirements of production deployment, and performance in a contended and continuous system environment, with a focus on extreme-scale I/O performance. Our research introduces a new framework for benchmarking of high-performance object stores for NWP, and significantly advances the understanding of the place of DAOS in the storage landscape for HPC systems as well as explores the features, configurations, and challenges such an object store presents for large-scale parallel I/O and for the design of I/O functionality for such applications.

\section{DAOS}

The Distributed Asynchronous Object Store (DAOS) is an open-source high-performance object store designed for massively distributed NVM. It provides a low-level key-value storage interface on top of which other higher-level APIs, also provided by DAOS, are built. Features include transactional non-blocking I/O, fine-grained I/O operations with zero-copy I/O to SCM, end-to-end data integrity, and advanced data protection. The OpenFabrics Interfaces (OFI) library is used for low-latency communications over a wide range of network back-ends.

DAOS is deployable as a set of I/O processes or engines, generally one per physical socket in a server node, each managing access to NVM devices associated with the socket. Graphical examples of how engines, storage devices and network connections can be arranged in a DAOS system, can be found at \cite{daos_architecture}.

An engine partitions the storage it manages into targets to optimize concurrency, each target being managed and exported by a dedicated group of threads. DAOS allows reserving space distributed across targets in \textit{pools}, a form of virtual storage. A pool can serve multiple transactional object stores called \textit{containers}, each with their own address space and transaction history.

Upon creation, objects are assigned a 128-bit unique identifier, of which 96 bits are user-managed. Objects can be configured for replication and striping across pool targets by specifying their \textit{object class}. An object configured with striping is stored by parts, distributed across targets, enabling concurrent access analogous to Lustre file striping.

\section{Weather field I/O}

NWP is used as an exemplar application as it is a HPC use case that has significant I/O requirements, that represent production I/O workloads, and has varied I/O patterns, enabling us to evaluate a range of I/O performance implications from these varied approaches. Additionally, due to the typically small size of data units being accessed, the application benefits from an object-like I/O approach to enable both efficient I/O operations from the production applications and easy data post-processing and querying for forecast generation, research, and analysis.

At the ECMWF, our exemplar weather forecasting centre, the NWP model is run 4 times a day in 1-hour time-critical windows using 2500 compute nodes. During each of these windows, 40 TiB of weather forecast data are generated and written by the model into the HPC storage system, and immediately read by post-processing tasks to generate derived products. 

%The NWP model under consideration in this paper is composed of a large set of parallel processes currently run across approximately 2500 HPC nodes. 
The data is generated by all model processes in a distributed manner, and it is sent through the low-latency interconnect to I/O servers, where it is aggregated. For the duration of a simulation, approximately 250 additional HPC nodes are devoted to running I/O server instances. After aggregation and data encoding, these servers forward the data to the HPC storage system.

Another 250 additional HPC nodes are devoted to running the post-processing tasks, after each step of the simulation, which read the generated output from the storage to then create derived data products.

In the described context, the I/O servers do not interact directly with the file system API. They do so via an intermediate domain-specific object store, the FDB5\cite{fdb-pasc19}, which is entirely software-defined. Such an object store saves any client software from implementing custom management of concurrent storage and indexing of weather fields in the file system, and enables object semantics for weather field access via a concise API.%, therefore removing software complexity.

The weather fields are the unit of data output by the model, and contain data for 2-dimensional slices covering the whole Earth surface for a given weather variable at a given time. They currently range between 1 and 5 MiB in size. Upon storage of a field, an indexing key must be provided together with the field data. A key is a set of field-specific key-value pairs that uniquely identify a field. Upon retrieval, the corresponding key has to be provided, and the field data is returned. %Fig. \ref{fig:nwp_schematic} shows a schematic representation of the described object store semantics.

%\begin{figure}[htbp]
%\centerline{\includegraphics[width=250pt]{introduction-object_storage_for_nwp-object_semantics.png}}
%\caption{Schematic representation of the weather field object store semantics implemented at the %centre.}
%\label{fig:nwp_schematic}
%\end{figure}

\subsection{Challenges}

With prospects of increasing simulation resolution and diversity of parameters modelled and output, %enabled by increases in computational power,
it is expected shortly the volume of forecast data generated will reach approximately 180 TiB per time-critical window, and up to 700 TiB per time-critical window in the near future.

Therefore, the pressure on the HPC storage system will increase significantly, requiring higher throughput to maintain the time-critical and operational aspects of the service. Exploration and adoption of novel storage technologies will be key for a satisfactory implementation of the upcoming resolution increases.

\subsection{Field I/O benchmark and object functionality}

A pair of C functions have been developed to perform writing and reading of weather fields to and from a DAOS object store, using the DAOS C API. They have been developed based on the design of FDB5, the domain-specific object store already employed within the ECMWF, so that the same type of operations as in operational workflows is carried out. The diagram in Fig. \ref{fig:fieldio_daos_design} shows the different DAOS concepts and APIs involved in the field I/O functions. 

\begin{figure}[htbp]
\centerline{\includegraphics[width=180pt]{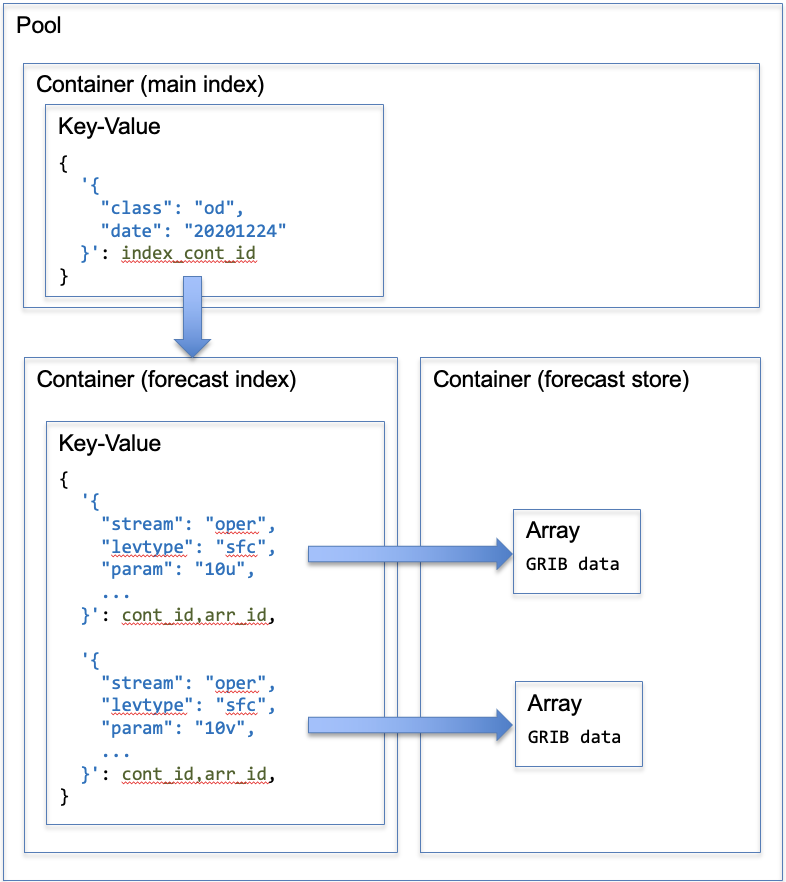}}
\caption{Diagram of DAOS concepts and APIs involved in the weather field I/O functions developed.}
\label{fig:fieldio_daos_design}
\end{figure}

At the top level a DAOS Key-Value object, in its own container, acts as a main index allowing location of the data belonging to a same model run or \textit{forecast}. That index maps the most-significant part of the key identifying a field (e.g. ``{`class': `od', `date': `20201224'}") to another DAOS container, at the lower layer. In that container is another Key-Value object that acts as a forecast index, enabling location of the data of the fields comprising a forecast. That index maps the least-significant part of the field key to a further DAOS container and a DAOS Array in that container where the data of the indexed weather field is stored.

How the developed functions perform field writes and reads using these concepts and objects is described, in a very simplified way, in Algorithms \ref{alg:algorithm1} and \ref{alg:algorithm2}, respectively.

\begin{algorithm}
\SetKwInput{KwData}{Inputs}
\caption{field write}
\KwData{field key, field data}
open main container\\
query most significant part of field key from Key-Value\\
\If{not found}
{
    create forecast index and store containers\\
    register forecast index container in main Key-Value\\
}
open forecast store container\\
write field data into new Array\\
open index container\\
update forecast Key-Value\\
\label{alg:algorithm1}
\end{algorithm}

\begin{algorithm}
\SetKwInput{KwData}{Inputs}
\SetKwInput{KwResult}{Outputs}
\caption{field read}
\KwData{field key}
\KwResult{field data}
open main container\\
query most significant part of field key from Key-Value\\
\If{not found}
{
    fail\\
}
open forecast index container\\
query least significant part of field key from Key-Value\\
\If{not found}
{
    fail\\
}
open store container\\
read field data from Array\\
\label{alg:algorithm2}
\end{algorithm}

When the field write function is called, with the binary field and its key as parameters, the most-significant part of the key is retrieved from the main Key-Value. If it exists, the indexed references to the forecast containers are retrieved and the containers are opened. If they do not exist, creation of a new pair of forecast index and store containers is attempted, with container IDs computed as md5 sums of the most-significant part of the key so that any concurrent processes attempting creation of the same pair of containers will avoid creation of inaccessible containers. Immediately after creation, the forecast containers are opened and the ID of the forecast store container is stored in a special entry in a newly created Key-Value in the forecast index container. Next, a reference to the forecast index container is indexed in the main Key-Value, using the most-significant part of the field key as key. Then, the binary field is written into an Array in the forecast store container, with a new object ID, and a reference to it is indexed in the forecast index Key-Value, using the least-significant part of the field key as key. %The field write is completed.

%As soon as a new field is written which does not share the most-significant part of its key with other existing fields in the store, a new pair of forecast containers, with an indexing Key-Value and a storing Array, respectively, are created. Besides, two fields with identical most-significant part of their key but different least-significant part, share a pair of forecast containers; in that case, the Key-Value in the forecast index container indexes the two least-significant parts to two different Arrays in the same forecast store container. 

Note that when a field is written with a key that already exists in the overall store, a new Array object is created and indexed, and the previously existing one is de-referenced. No read-modify-write is performed upon re-write, and the functions do not delete de-referenced objects by design.

When the field read function is called, with the key of a field the parameter, the most-significant part of the key is looked up in the main Key-Value. If it exists, the indexed reference to the forecast containers are retrieved and the containers are opened. Next, the least-significant part of the key is retrieved from in the forecast index Key-Value. If that is present, the indexed references to the forecast store container and Array are retrieved. Finally, the forecast store container is opened and the Array is read. %The field read is completed.

For the NWP applications we are investigating, the separation of the different Key-Values and Arrays in different containers is motivated by the need to avoid contention for the same container or indexing Key-Value object during intensive I/O workloads. This separation also allows the implementation of the different parts of the object store with different storage back-ends or on different systems.

%To test the proper functioning and consistency of the functions, DAOS has been installed in a virtual machine and deployed with a single server and two engines, using the file system to emulate NVMe and main memory to emulate SCM. The functions have been run in I/O-intensive workloads, and checks have been made to ensure the written values are returned without corruptions or failures.

\section{Methodology}

To assess the performance of DAOS, different I/O workloads have been generated and run in a research HPC system using the well-known IOR\cite{ior_repo} benchmark and the field I/O functions. The performance of DAOS and the benchmarks in the HPC has been analysed based on different throughput definitions discussed in subsequent sections.

\subsection{IOR}

IOR is a community-developed I/O benchmark which relies on MPI to run and coordinate parallel processes performing I/O operations against a storage system. It includes back-ends to operate with various popular storage systems, including DAOS.

IOR has been employed in this analysis with the intent to measure the throughput an application could achieve if it were programmed to operate in a traditional parallel, bulk synchronous I/O manner. That is, running a number of parallel client processes which use the high-level DAOS Array API to write or read data from a DAOS object store, all of them starting each I/O operation simultaneously, and waiting for each other to finish.

For the tests in this analysis, the IOR benchmark has been run in \textit{\textbf{segments}} mode. IOR is invoked with a set of parameter values which instruct each client process to perform a single I/O operation, transferring its full data size. This is with the intent to assess the performance of a hypothetical optimised parallel application which is designed to minimise the number of I/O operations interacting with the storage. Processes in such applications issue a single transfer operation comprising all the data parts they manage, in contrast to an equivalent, non-optimised application where processes issue a transfer or even an open and a close operation for each data part.

Unless the storage is not optimised to handle large transfers or objects, this benchmark mode should give an idea of what is the maximum, ideal throughput the storage can deliver. This mode has been implemented by setting both the \verb!-b! and \verb!-t! IOR parameters to the size of each data part managed by each process, and \verb!-s! to the number of data parts managed by each process. The \verb!-i! parameter has been set to 1, and the \verb!-F! flag (file per process) has been enabled. Each process performs the following during the benchmark execution: a) initial barrier, b) pre-I/O barrier, c) object create/open of size \verb!t! * \verb!s! bytes, d) transfer (write or read) of \verb!t! * \verb!s! bytes, e) object close, f) post-I/O barrier, g) post-I/O processing and logging, h) final barrier.

\subsection{Field I/O}

To gain insight on how DAOS performs in the previously outlined NWP I/O server use case, we use our custom field I/O benchmark, which mimics operational NWP I/O workflows. Here, parallel processes perform a sequence of field I/O operations using the functions previously described in the Weather field I/O section, with no synchronisation. Pool and container connections in a process are cached once initialised.

In contrast to the IOR benchmark in segments mode, the processes in this benchmark perform multiple I/O operations of smaller sizes, one for each data part. Each field I/O operation involves an Array open, transfer and close, and usually involves a few operations with Key-Value objects. We also do not enforce I/O synchronisation, as in IOR.%, and there is no intermediate processing between I/O operations.

Fig. \ref{fig:benchmark_config} shows a schematic representation of the processes and sequence of field I/O operations involved in an example benchmark run.

\begin{figure}[htbp]
\centerline{\includegraphics[width=250pt]{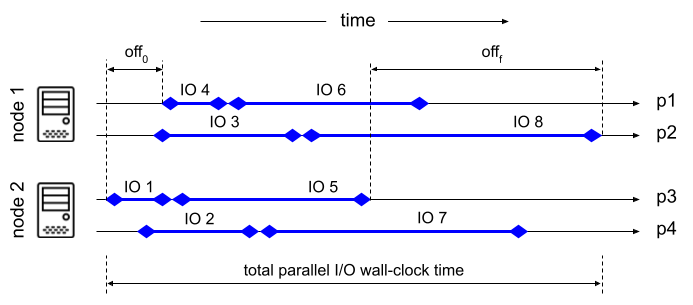}}
\caption{Processes and I/O operations involved in an example Field I/O benchmark run.}
\label{fig:benchmark_config}
\end{figure}

The field I/O benchmark has been configured in three different modes for this evaluation:

\begin{itemize}[leftmargin=*]
    \item \textit{\textbf{full}}: the fully-functional field I/O functions are used as described in Weather field I/O.
    \item \textit{\textbf{no containers}}: the use of container layers in the field I/O functions is disabled with the intent to analyse any potential issues in the use of container layers. The functions create and operate with all DAOS Key-Values and Array objects in the main container.
    \item \textit{\textbf{no index}}: the use of indexing Key-Value objects and container layers in the field I/O functions are disabled with the intent to compare with results from other benchmark modes and assess the overhead of field indexing. To preserve the ability to locate previously written fields, the field I/O functions map the field identifiers to DAOS Array object IDs by calculating a md5 sum of the identifiers. The arrays are stored and read directly from the main container.
\end{itemize}

By default, in modes with indexing enabled, each parallel process writes and reads fields indexed in its own forecast index Key-Value and therefore there should be little contention for the different objects within DAOS. The benchmark, however, can be configured to have a single shared forecast index Key-Value among all processes, inducing high contention on that Key-Value object.

The Field I/O benchmark is maintained and documented in a publicly accessible open-source Git repository\cite{field_io_repo}.

\subsection{Access patterns}

The IOR and field I/O benchmarks, in each of their modes, can be further adjusted and employed as part of larger workflows to generate specific combinations of I/O workload of interest, hereafter referred to as \textit{access patterns}. Two access patterns used in this analysis are described next.

\begin{itemize}[leftmargin=*]
    \item \textit{\textbf{A (unique writes then unique reads)}}: This access pattern has two phases. %In the first phase a number of client processes is executed in each of the client nodes employed for the benchmark run.
    Each client process performs a number of writes to new objects in the storage (single transfer to single object for IOR in segments mode). Once all writer processes have finished, a second phase begins, where another process set of the same size and distribution is executed performing an equivalent number of reads from the objects generated in the write phase. The access pattern ends when all second-phase processes terminate.
    
    This access pattern is designed to provide a situation where there is no contention for same fields and when there is only either write or read workloads in operation at any one time, analogous to a single large-scale application utilising the object store, or to systems where the predominate I/O patterns are only one I/O phase (writing or reading).
    
    In this access pattern, when run with the IOR benchmark, there is no contention in Array writes or reads, as each process accesses its own Array independently from others. When run with the field I/O benchmark in any of its modes with a forecast index Key-Value per process, there is no contention in any of the forecast index or store objects, as each process writes or reads its own Key-Value and Arrays sequentially.
    
    \item \textit{\textbf{B (repeated writes while repeated reads)}}: This access pattern starts with a setup phase to populate the storage with data, where half of the client processes (and thereby half the client nodes) perform a single write to a new object. Once all writer processes have terminated, the main phase begins. Half the client processes perform a number of re-writes to designated objects; simultaneously, the other half of the client processes perform the same number of reads from their designated objects. The access pattern ends when all main-phase processes terminate.

    This access pattern has only been implemented with the field I/O benchmark as IOR does not allow coordination between a set of writers and a set of readers. It aims to mimic situations where one or more large-scale applications issue simultaneous writes and reads of the same objects or fields. This is the equivalent of the I/O behaviour of actual NWP and product generation workloads, or to systems with mixed applications of varying I/O workloads.
    
    In this access pattern, when run with Field I/O in \textit{full} or \textit{no containers} modes with a forecast index Key-Value per process, there is no contention in Array writes or reads, but there is some contention in each forecast index Key-Value between reader and writer processes on the same object. When run in \textit{no index} mode, the same degree of contention occurs at the Array level.
\end{itemize}

%The number of processes per client node and I/O iterations per process can be adjusted in both access patterns.

\subsection{Parameter variation}

The access patterns presented can be configured not only to run with the different benchmarks and modes, but also to use a range of parameter values. The following list summarises these parameters and the values tested in this paper.

\begin{itemize}
    \item \textit{\textbf{number of DAOS server nodes}}: 1, 2, 4, 8, 10, 12, 14, 16
    \item \textit{\textbf{number of client nodes}}: 0.5x, 1x, 2x and 4x the amount of server nodes
    \item \textit{\textbf{number of processes per client node}}: 1, 3, 4, 6, 8, 9, 12, 18, 24, 36, 48, 72, 96
    \item \textit{\textbf{number of iterations or data parts per proc.}}: 100, 2000
    \item \textit{\textbf{object class}}: \verb!S1! (no striping), \verb!S2! (striping across two targets) and \verb!SX! (striping across all targets)
    \item \textit{\textbf{object size}}: 1, 5, 10, 20 and 50 MiB
\end{itemize}

\subsection{Throughput definitions}

To quantify, compare, and discuss the performance delivered by the storage, a set of metric and throughput definitions are introduced here and applied in the Results section. During execution, the IOR and Field I/O benchmarks report timestamps for various events. Each timestamp is reported together with an identifier of the client node, process and iteration or I/O they belong to, as well as the event name. We record timestamps for: execution start, I/O start, object open start, object open end, data write/read start, data write/read end, object close start, object close end, I/O end, and execution end.

In Field I/O, \textit{I/O start} is recorded immediately before calling the field write or read functions. In IOR it is equivalent to \textit{object open start}.

The following derived metrics are calculated where relevant from the timestamps.

\begin{itemize}[leftmargin=*]
    \item \textit{\textbf{single iteration parallel I/O wall-clock time}}: calculated as the maximum I/O end minus the minimum I/O start for a given I/O iteration across all parallel processes. This metric is only valid in benchmarks where I/O is synchronised.
    \item \textit{\textbf{total parallel I/O wall-clock time}}: calculated as the maximum I/O end of the last I/O iteration minus the minimum I/O start of the first I/O iteration across all parallel processes. This metric is calculated separately for the write and read phases of the described access patterns, and is valid for synchronised and non-synchronised benchmarks. In the example in Fig. \ref{fig:benchmark_config}, the metric would be calculated as $t_f(IO8) - t_0(IO1)$.
\end{itemize}

With these derived metrics, the throughput is calculated in two different ways, outlined below. Information such as I/O size and number of parallel processes is available at analysis time and is used as input for the calculation of these throughputs.

\begin{itemize}[leftmargin=*]
    \item \textit{\textbf{synchronous bandwidth}}: the sum of the sizes of the I/O operations for an iteration across all parallel processes, divided by the single iteration parallel I/O wall-clock time. The averaged result across all iterations is the synchronous bandwidth, only applicable to synchronous benchmarks (i.e. IOR).
    \begin{equation*}
    \begin{split}
    sync.\ bw. = \frac{1}{n} * \sum_{i=1}^{n} \frac{\sum_{j=1}^{m} size(IO_{ij})}{max_j\{t_f(IO_{ij})\} - min_j\{t_0(IO_{ij})\}} \\
    n = \#iterations,\ m = \#processes\label{eq1}
    \end{split}
    \end{equation*}
    % This definition is only valid for the synchronised IOR benchmark.
    IOR internally calculates the synchronous bandwidth following this definition.

    \item \textit{\textbf{global timing bandwidth}}: the sum of the I/O sizes across all I/O operations in the workload and divided by the total parallel I/O wall-clock time.
    \begin{equation*}
    \begin{split}
    \ \ global\ timing\ bw. = \frac{\sum_{i=1}^{n} size(IO_i)}{t_f(IO_n) - t_0(IO_1)} \\
    n = \#I/O\ operations\label{eq2}
    \end{split}
    \end{equation*}
    It is calculated separately for the write and read phases of the described access patterns.
    The value of this bandwidth is inversely proportional to global parallel I/O wall-clock time. If a benchmark performs any work between I/O iterations this will impact the global timing bandwidth measure as it will increase the total parallel I/O wall-clock time.
\end{itemize}

\subsection{Asynchronous I/O distribution}

When running patterns A or B with Field I/O, the distribution of I/O operations should be analysed and taken into account at benchmark configuration time to ensure the resulting distribution is representative of the desired access pattern. To characterise the overall distribution of I/Os, the following offsets can be measured for a benchmark run:

\begin{itemize}[leftmargin=*]
    \item \textit{\textbf{initial I/O offset ($off_0$)}}: measured separately for the write and read phases of an access pattern, calculated as the last first-iteration I/O start time minus the first I/O start time across all processes, as shown in Fig. \ref{fig:benchmark_config}.
    \item \textit{\textbf{final I/O offset ($off_f$)}}: measured separately for the write and read phases of an access pattern, calculated as the last I/O end time minus the first last-iteration I/O end time.
    \item \textit{\textbf{phase start offset ($po_0$)}}: calculated as the absolute value of the difference between the first I/O start time of the write and read phases, only applicable to access pattern B.
    \item \textit{\textbf{phase end offset ($po_f$)}}: calculated as the absolute value of the difference between the last I/O end time of the write and read phases, only applicable to access pattern B.
\end{itemize}

For the most accurate results, we should ensure $off_0$ and $po_0$ are not large relative to the total parallel I/O wall-clock time by adjusting access pattern configuration.

$off_f$ may account for a large portion of the total parallel I/O wall-clock time as, for example, the storage may process operations from the various parallel processes in an unbalanced manner. Similarly, $po_f$ may account for a large portion of the total duration of I/O (write and read) of a pattern B run, for example, as the storage may process write and read operations in an unbalanced manner.

Outside these offset time spans, there are as many in-flight I/O operations as parallel processes at any point in time during the benchmark run.

\section{Results}

Results obtained are discussed next, including selected system and DAOS configurations, and bandwidth calculations for the different benchmarks and access patterns.

\subsection{Experiment environment}

The benchmarks we present have been conducted using the NEXTGenIO system\cite{ngio}, which is a research HPC system composed of dual-socket nodes with Intel Xeon Cascade Lake processors. Each socket has six 256 GiB first-generation Intel's Optane DCPMMs configured in AppDirect interleaved mode. Each processor is connected with its own network adapter to a low-latency OmniPath fabric. Each of these adapters has a maximum bandwidth of 12.5 GiB/s. The fabric is configured in dual-rail mode, that is, two separate OmniPath switches interconnect first-socket adapters and second-socket adapters separately, respectively.

The HPC system nodes use CentOS7 as the operating system, with DAOS v2.0.1. For the majority of the tests, two DAOS engines have been deployed in each node used as DAOS server, one on each socket using the associated fabric interface and interleaved SCM devices, with 12 targets per engine. %SCM devices in the client nodes have not been used.

\subsubsection{Fabric provider}
The OFI PSM2 fabric provider implements Remote Direct Memory Access (RDMA) over an OmniPath fabric, of particular interest in the HPC system employed. However, use of PSM2 in DAOS is not yet production-ready, impeding dual-engine per node, dual-rail DAOS deployments.

Due to this, OFI's TCP provider has been used for the majority of the test runs in the analysis. TCP's use of operating system sockets instead of RDMA has an impact in transfer bandwidth between nodes, which is further detailed in IOR results and in the High-performance network section. %This sets expectations of what the maximum achievable application bandwidth is when using the TCP provider.

\subsubsection{Process pinning}
The process pinning strategy has been found to have substantial impact in I/O performance, with up to 50\% reduction in performance when DAOS server engines were not optimally pinned across processors in a node, and up to 90\% reduction when benchmark client processes were not optimally pinned. For best performance, each DAOS engine has been configured to pin engine processes to a single socket, and target the attached fabric interface. This has been achieved via the \verb!pinned_numa_node! DAOS server configuration item. On the client side, processes have been distributed in a balanced way across sockets in each client node.

\subsection{IOR results}

In this test set, the IOR benchmark in segments mode has been run (access pattern A) with the intent to analyse the maximum write and read performance a client application can achieve.

In our initial experiment we deployed DAOS on a single server node, with one or two engines, each running on a socket of the node and using the corresponding fabric adapter (\verb'ib0' and \verb'ib1', respectively). The IOR client processes have been pinned to one or two sockets (or interfaces) in one or two client nodes, as detailed in Table I. The segment count has been set to 100.% after testing a few other values and finding that the bandwidth metrics do not show large variability with this segment count. 

A segment size of 1 MiB has been motivated by the current weather field size at the exemplar weather centre. This segment size results in objects of 100 MiB in size being written or read by IOR. Striping has been disabled (\verb!OC_S1!) to avoid complexity in network behaviour for the initial tests. The number of processes per client node has been set to 24, 48, 72 and 96 after verifying with several other IOR tests that process counts in that range usually resulted in higher benchmark bandwidths. Each test has been repeated 9 times for the range of client processes, and the maximum synchronous bandwidth obtained among the 36 repetitions is reported in Table I for both the write and read phases.

\begin{table}[htbp]
\caption{Access Pattern A, IOR Segments, 1 Server Node}
\begin{center}
\begin{tabular}{|c|c|c|c|c|}
\hline
\textbf{server}&\textbf{engines per}&\textbf{ifaces per}&\multicolumn{2}{|c|}{\textbf{bandwidth (GiB/s)}} \\
\cline{4-5} 
\textbf{nodes}&\textbf{server node}&\textbf{client node}& \textbf{\textit{1 client node}}& \textbf{\textit{2 client nodes}} \\
\hline
1 & 1 (ib0) & 1 (ib0) & 3.0w / 4.2r & 2.6w / 6.2r  \\
\hline
1 & 1 (ib0) & 2 & 3.0w / 7.4r & 2.9w / 7.7r  \\
\hline
%1 & 1 (ib1) & 1 (ib1) & 1.5w / 4.0r & 1.2w / 4.0r  \\
%\hline
%1 & 1 (ib1) & 2 & 1.5w / 3.8r & 1.5w / 2.9r  \\
%\hline
1 & 2 & 2 & 5.5w / 7.5r & 5.5w / 9.5r  \\
\hline
\end{tabular}
\label{tab1}
\end{center}
\end{table}

With a single server engine using one interface and a set of client processes using a single client interface on the first socket of the corresponding nodes (first row of the table; 1 client node), the maximum write bandwidth reaches 3 GiB/s and, for read, 4.2 GiB/s. When using multiple client sockets and interfaces against a single server engine on the first socket and interface (last column of first row and second row), the maximum write bandwidth still saturates at 3 GiB/s whereas the maximum read bandwidth reaches up to 7.7 GiB/s.

%When running the servers and clients on the second socket with the corresponding interfaces (third row of the table), the maximum bandwidth drops to 1.5 GiB/s for write and 4.0 GiB/s for read, possibly due to a network imbalance detected with other tests out of the scope of this report.

For the test in the last row of the table, with two DAOS engines deployed in a server node, and using both interfaces on a single client node (last row; 1 client node), the maximum write bandwidth reaches up to 5.5 GiB/s. For read, the maximum bandwidth remains at 7.5 GiB/s on a single client node, whereas it reaches up to 9.5 GiB/s with two client nodes (a total of 4 network interfaces). This indicates that more client interfaces than server interfaces are necessary to saturate server interface bandwidth for read, which is likely to be the deployment configuration of DAOS for production systems.

To further analyse the bandwidth increase when more engines and interfaces are employed, the same access pattern with identical parameters has been run with different numbers of server and client nodes, using both processors and network interfaces on each node. Using as many client nodes as server nodes has been pursued for all tested configurations to effectively make use of the available bandwidth theoretically provided by the server interfaces, and, where possible, configurations with twice or four times as many client nodes have been tested to explore if, as suggested by results in Table I, these result in higher bandwidths. A few special configurations have also been tested, for instance with less client nodes than server nodes (e.g. 2 server nodes and 1 client node), or with slightly less than double the server nodes (e.g. 10 server nodes and 18 client nodes) to explore bandwidth behaviour in unbalanced configurations. Fig. \ref{fig:ior} shows, for each combination of server and client node count, the mean synchronous bandwidth obtained across all repetitions for the best performing number of client processes per client node.

\begin{figure}[htbp]
    \centering
    \begin{subfigure}[b]{200pt}
        \centering
        \includegraphics[width=200pt]{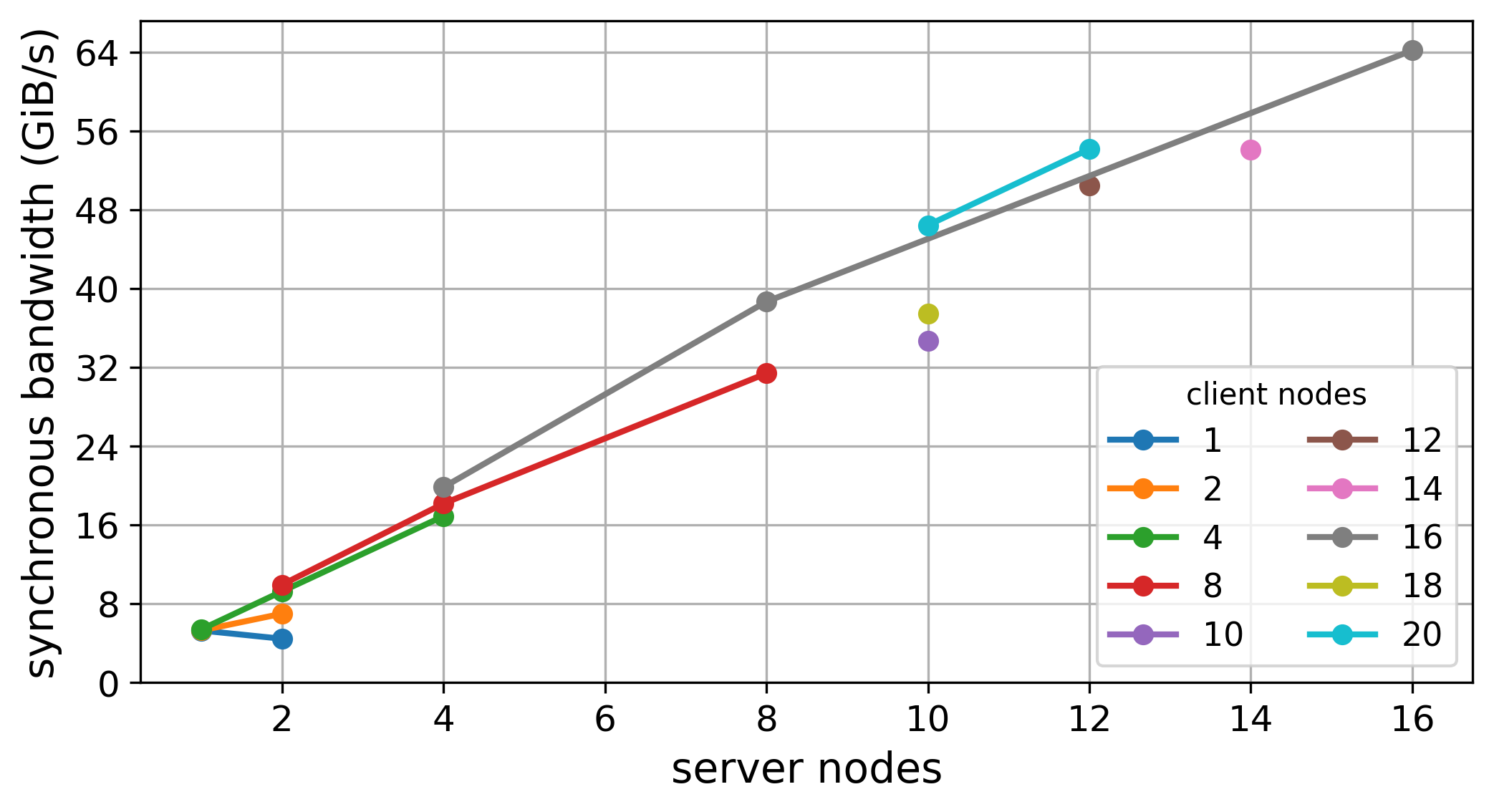}
        \caption{Write}
    \end{subfigure}
    %\vskip\baselineskip
    \begin{subfigure}[b]{200pt}
        \centering
        \includegraphics[width=200pt]{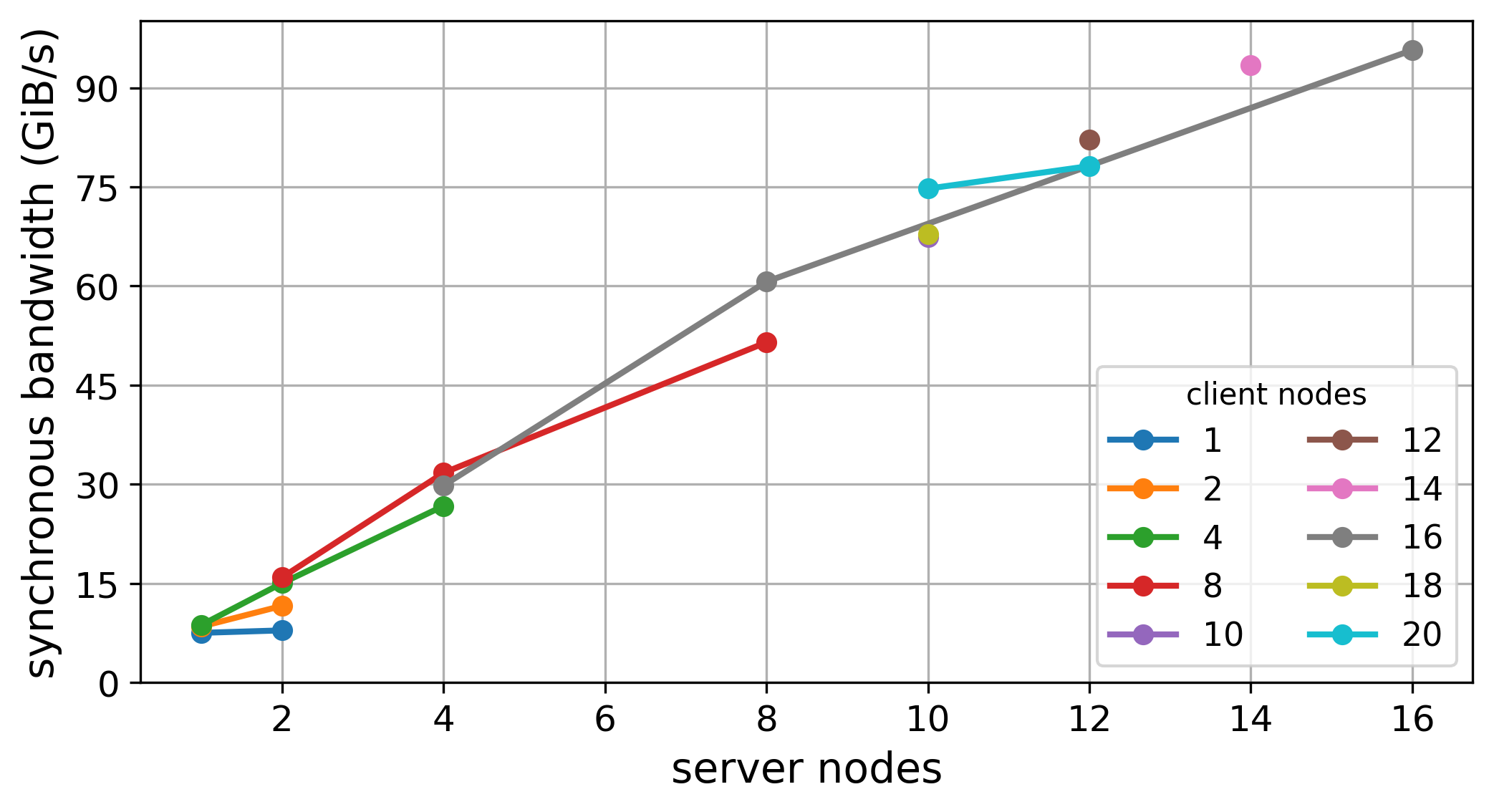}
        \caption{Read}
    \end{subfigure}
    \caption{Mean synchronous write and read bandwidth results for access pattern A (unique writes then unique reads) with IOR DAOS in segments mode.}
    \label{fig:ior}
\end{figure}

For a single dual-engine server, the bandwidth achieved per engine is of approximately 2.5 GiB/s for write and 5 GiB/s for read. As the number of servers increases, the bandwidths increase linearly at a rate of approximately 2.5 GiB/s write and 3.75 GiB/s read for every additional engine.%, showing an encouraging linear increase in bandwidth with additional server nodes. 

It is worth noting that configurations with twice as many client nodes generally deliver best performance, both for write and for read. Setups with a substantially higher ratio (e.g. four times as many) of client nodes do not show significant increases in performance, and configurations with a slightly lower ratio show a substantial decrease in performance.

Above 8 server nodes, the scaling rate seems to decrease slightly even if using twice as many client nodes, as seen in the bandwidth results for the setup with 10 server nodes and 20 client nodes.

The fact that the increase in read bandwidth per engine is not equal to the bandwidth achieved with a single engine is possibly due to the fact that multiple engines on both sockets in different nodes may contend to communicate through a single interface on one socket. Nonetheless, the linear increase in bandwidth with increasing engines is encouraging.

\subsection{High-performance network}

Due to the limitations encountered with the PSM2 fabric provider, we have not been able to benchmark DAOS with the highest-performance network functionality. However, it is possible to deploy DAOS with PSM2 using a single engine per server and only one socket used per client node. This restricts the usable network, processing, memory, and storage resources within nodes, but does allow comparison of TCP and PSM2 configurations of DAOS and an assessment of the impact of using a lower-performance network configuration.

We benchmarked a configuration using 4 DAOS server nodes, and up to 16 client nodes, which we tested with a number of different process counts (4, 8, 12, and 24 processes per client node). We used IOR in segments mode with the same configuration as used previously.

Fig. \ref{fig:psm2_vs_tcp} compares both read and write performance when using the IOR benchmark with the different communication mechanisms. It is evident that PSM2 does provide higher performance than TCP, especially when scaling up the number of processes per client node. PSM2 demonstrates 10\% to 25\% higher bandwidth than TCP, and can also provide higher performance at lower node counts. However, it can be observed that the DAOS performance using PSM2 follows the same general patterns as with TCP communications and, whereas we expect the former would result in larger absolute bandwidths by up to a 25\%, we consider their behaviour at scale would be similar.

%Therefore, given that use of TCP enables a wide range of DAOS engines and client processes to be assessed (using the full resources of nodes), and larger storage capacity to be exploited, we evaluate that the TCP approach used for this study enables a reasonable evaluation of DAOS to be undertaken.

\begin{figure*}[htbp]
    \centering
    \begin{subfigure}[b]{125pt}
        \includegraphics[width=125pt]{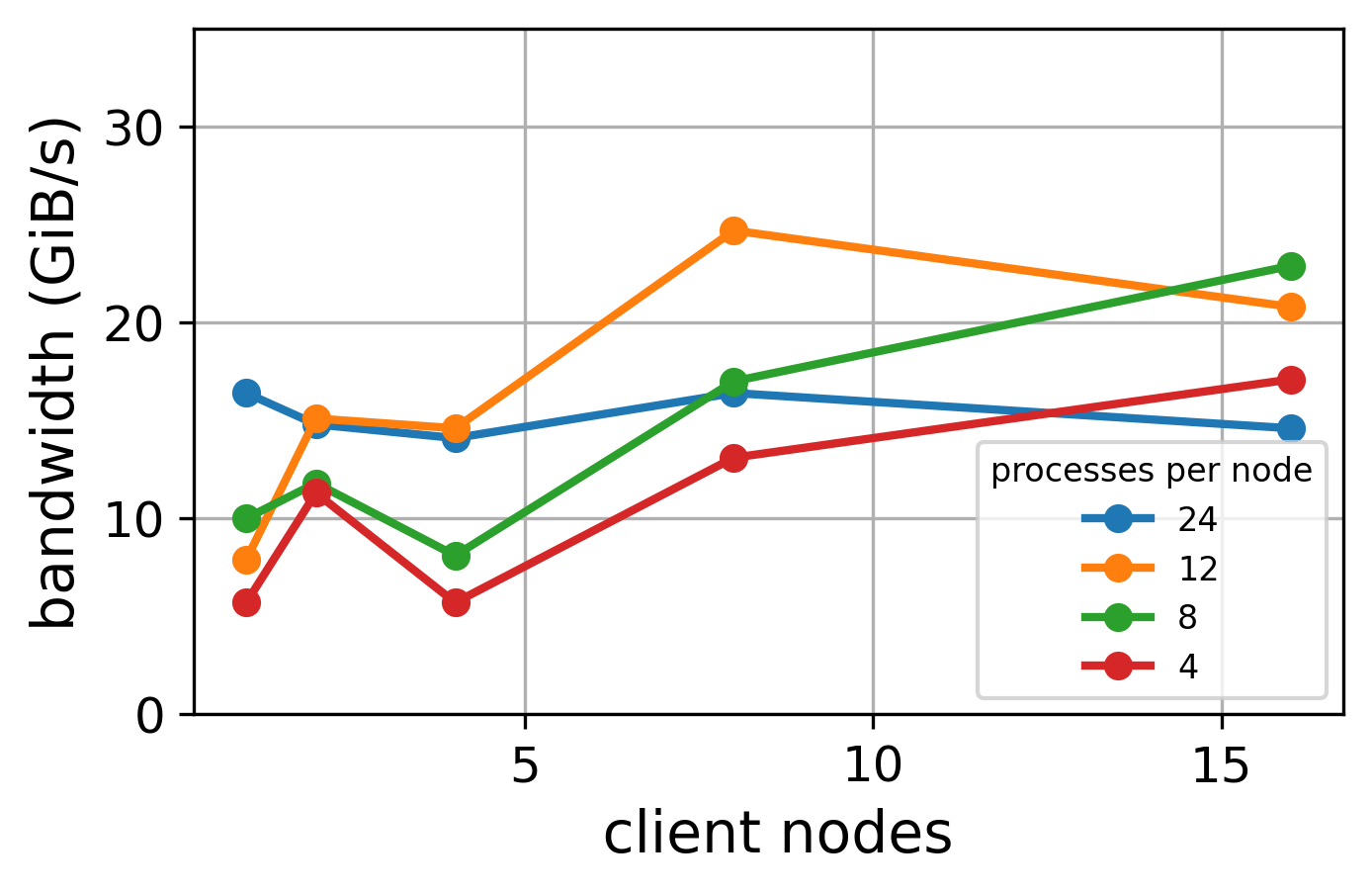}
        \caption{PSM2 networking, write}
    \end{subfigure}
    \begin{subfigure}[b]{125pt}
        \includegraphics[width=125pt]{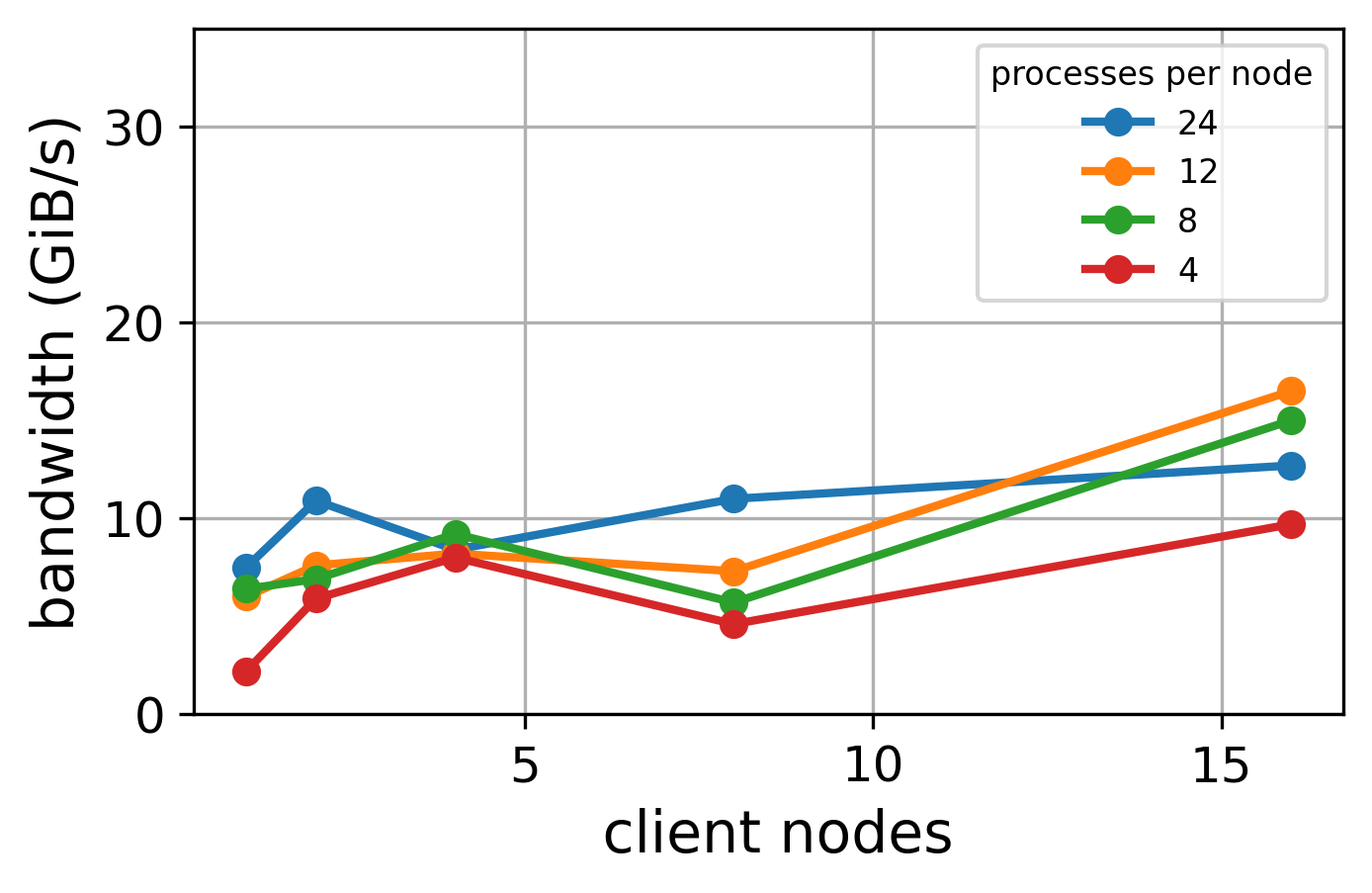}
        \caption{TCP networking, write}
    \end{subfigure}
    \begin{subfigure}[b]{125pt}
        \includegraphics[width=125pt]{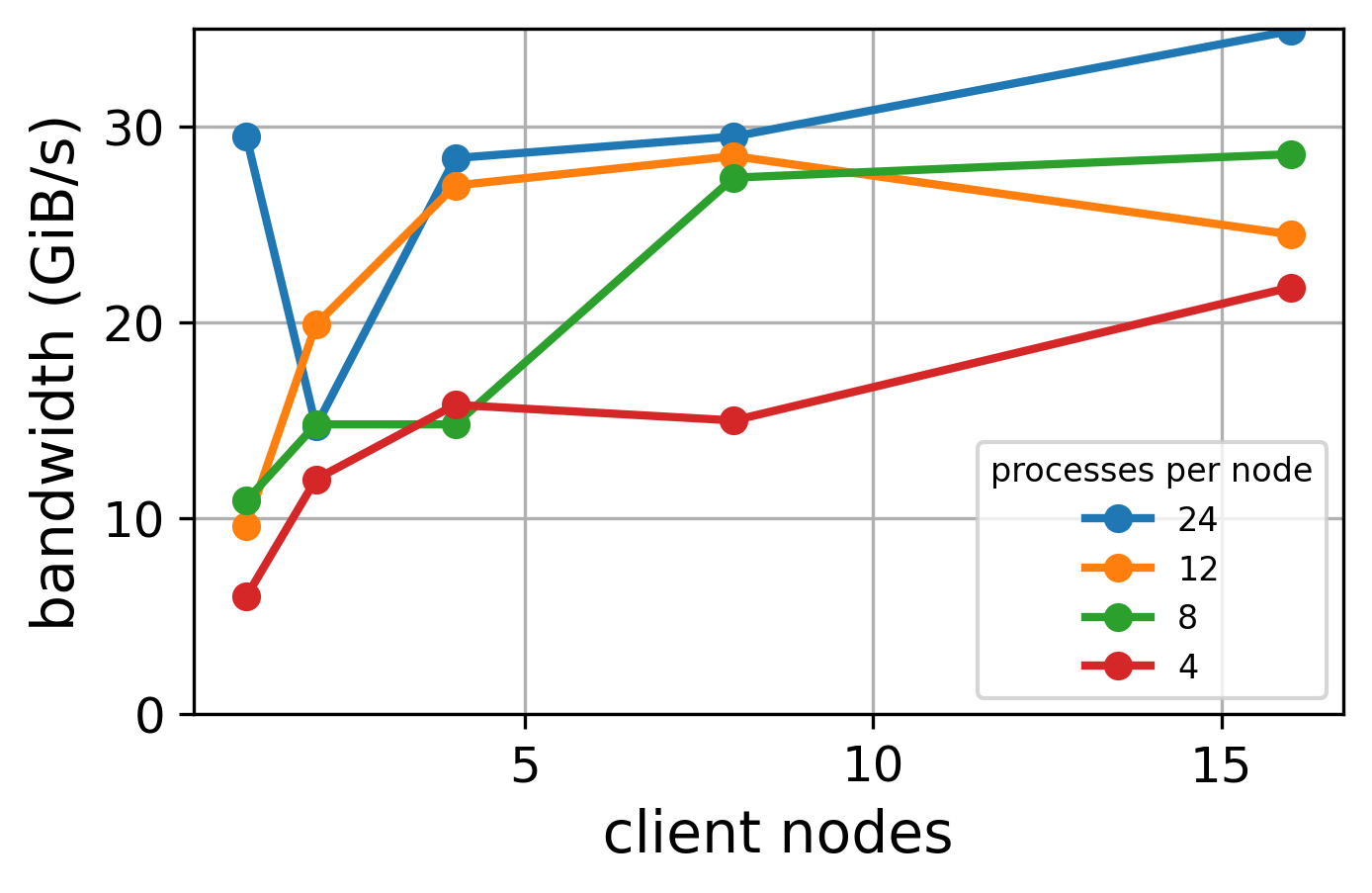}
        \caption{PSM2 networking, read}
    \end{subfigure}
    \begin{subfigure}[b]{125pt}
        \includegraphics[width=125pt]{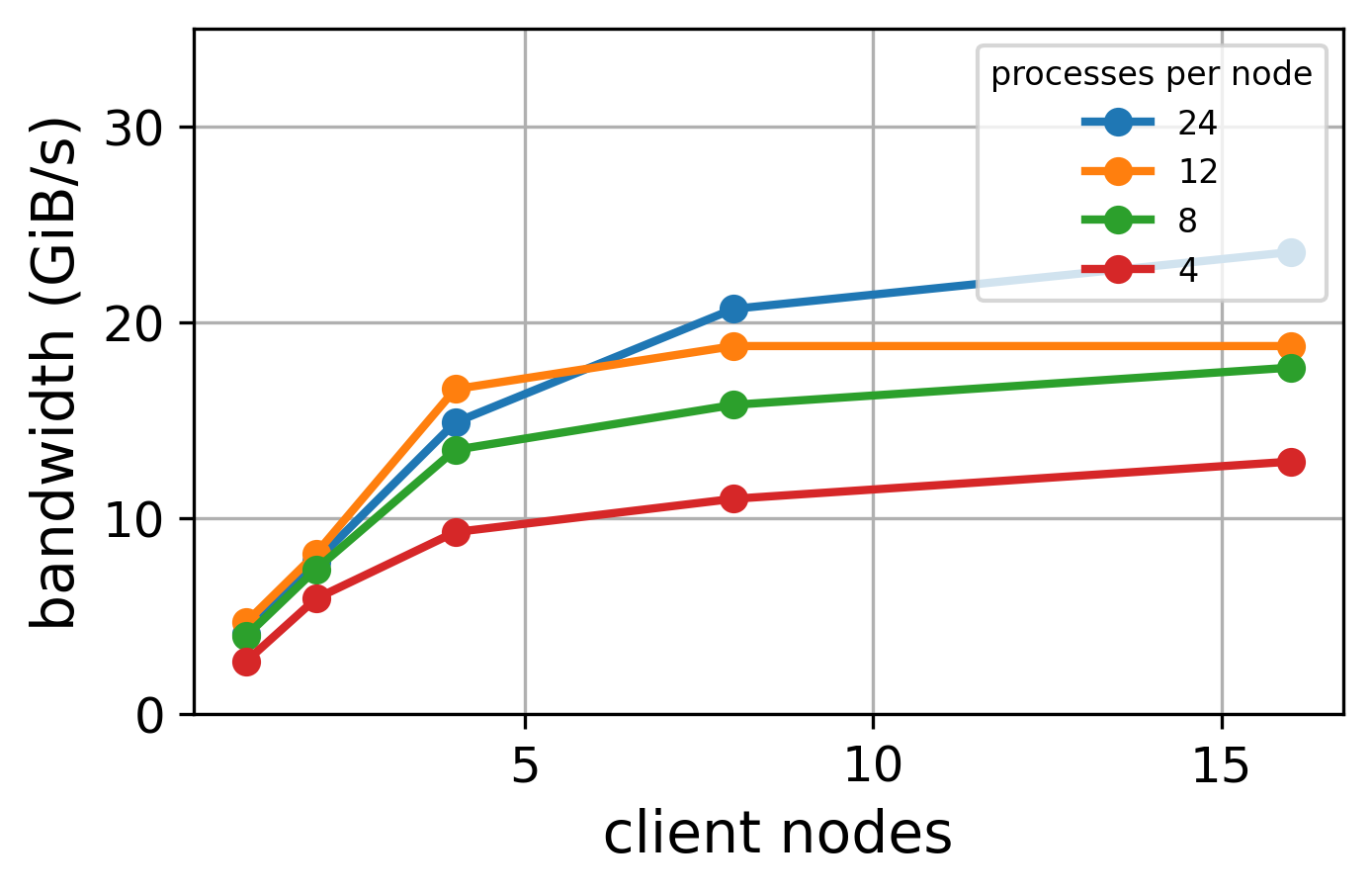}
        \caption{TCP networking, read}
    \end{subfigure}
    %\vskip\baselineskip
    \caption{IOR segments benchmark using 4 DAOS server nodes and a range of IOR client nodes, comparing TCP vs PSM2 network.}% The bandwidth scales well, and there is no performance degradation in pattern B.}
    \label{fig:psm2_vs_tcp}
\end{figure*}

%, and we consider that it does not change the performance achievable by DAOS on this platform significantly enough for the TCP benchmarking approach to be invalid or material different to the performance when using a more optimised communication mechanism. 

\subsection{Field I/O results}

We ran the Field I/O benchmark with the intent to analyse the order of magnitude and behaviour of bandwidths obtained with the field I/O functions rather than simple segment transfers; how the different field I/O modes scale; and how bandwidth behaves in access pattern B with an I/O workload similar to operational workloads.

For the rest of the tests in this paper, both sockets and network interfaces have been used on each of the nodes employed for the benchmark runs, with two DAOS engines deployed on each server node and client processes pinned to both sockets on each client node.

\subsubsection{Selection of client node and process count}

As a first parameter exploration and selection exercise, and with the intent to reduce the test parameter domain, we have run access pattern A with Field I/O, employing two server nodes and different amounts of client nodes (1, 2, 4 and 8) and processes counts (values listed in Parameter variation).

The number of I/O operations per client process has been set to 2000 to reduce the effect of any process start-up delays in bandwidth measurements. The Array object size has been set to 1 MiB, again motivated by the NWP use case.

The Key-Value objects have been configured with striping across all targets (\verb!OC_SX!), and no striping (\verb!OC_S1!) has been configured for the Array objects. This configuration was chosen on the assumption that Key-Values can be accessed by multiple process at a time for these benchmarks and could therefore benefit from striping across targets, whereas Arrays are never accessed by more than one process simultaneously. % We evaluate whether this assumption holds later in this section.

Benchmark runs have been repeated 10 times. The average global timing bandwidths are shown in Fig. \ref{fig:fieldio_cn_cpcn}.

\begin{figure}[htbp]
    \begin{subfigure}[b]{120pt}
        \includegraphics[width=120pt]{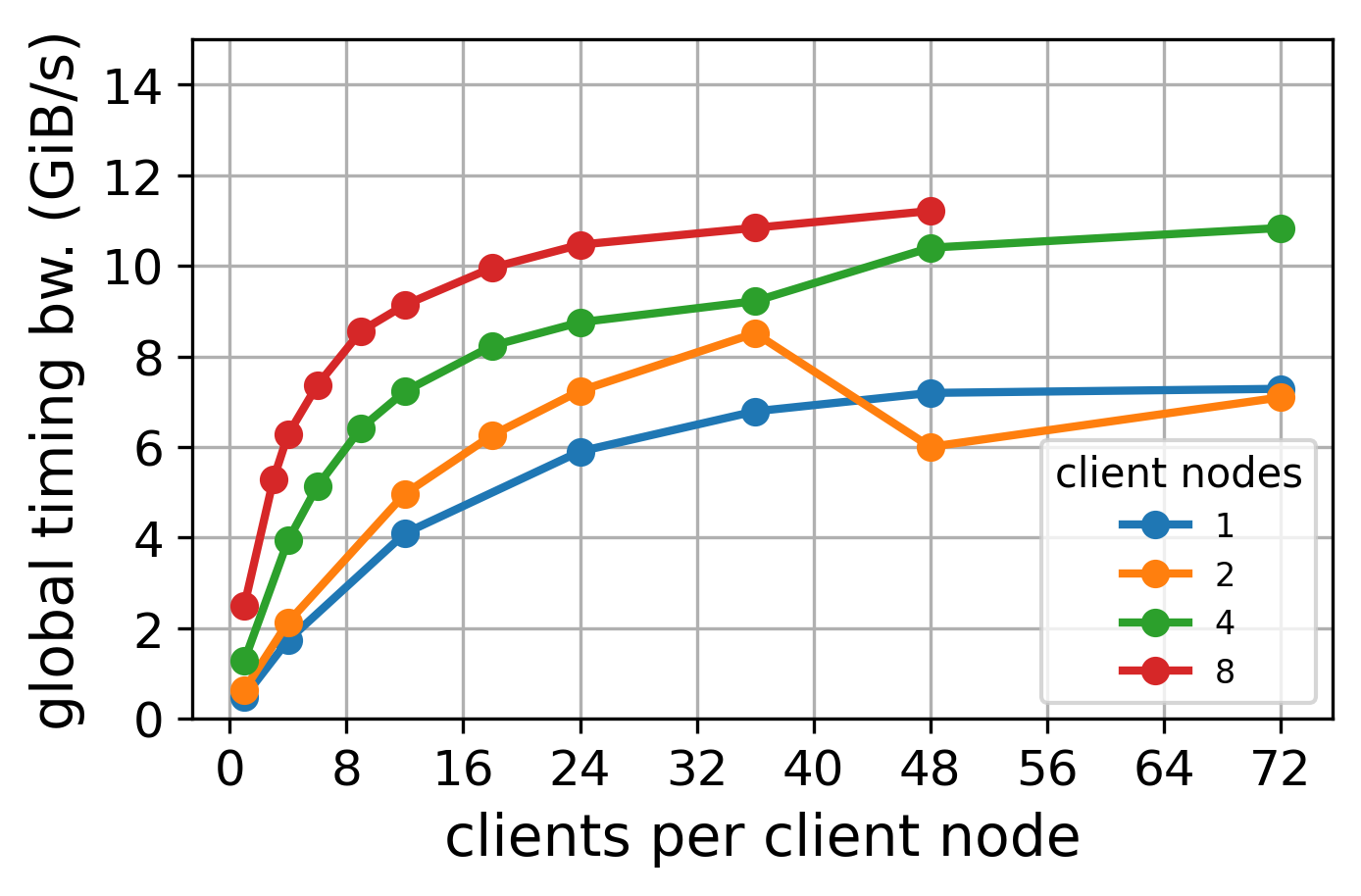}
        \caption{Access pattern A, write}
    \end{subfigure}
    \begin{subfigure}[b]{120pt}
        \includegraphics[width=120pt]{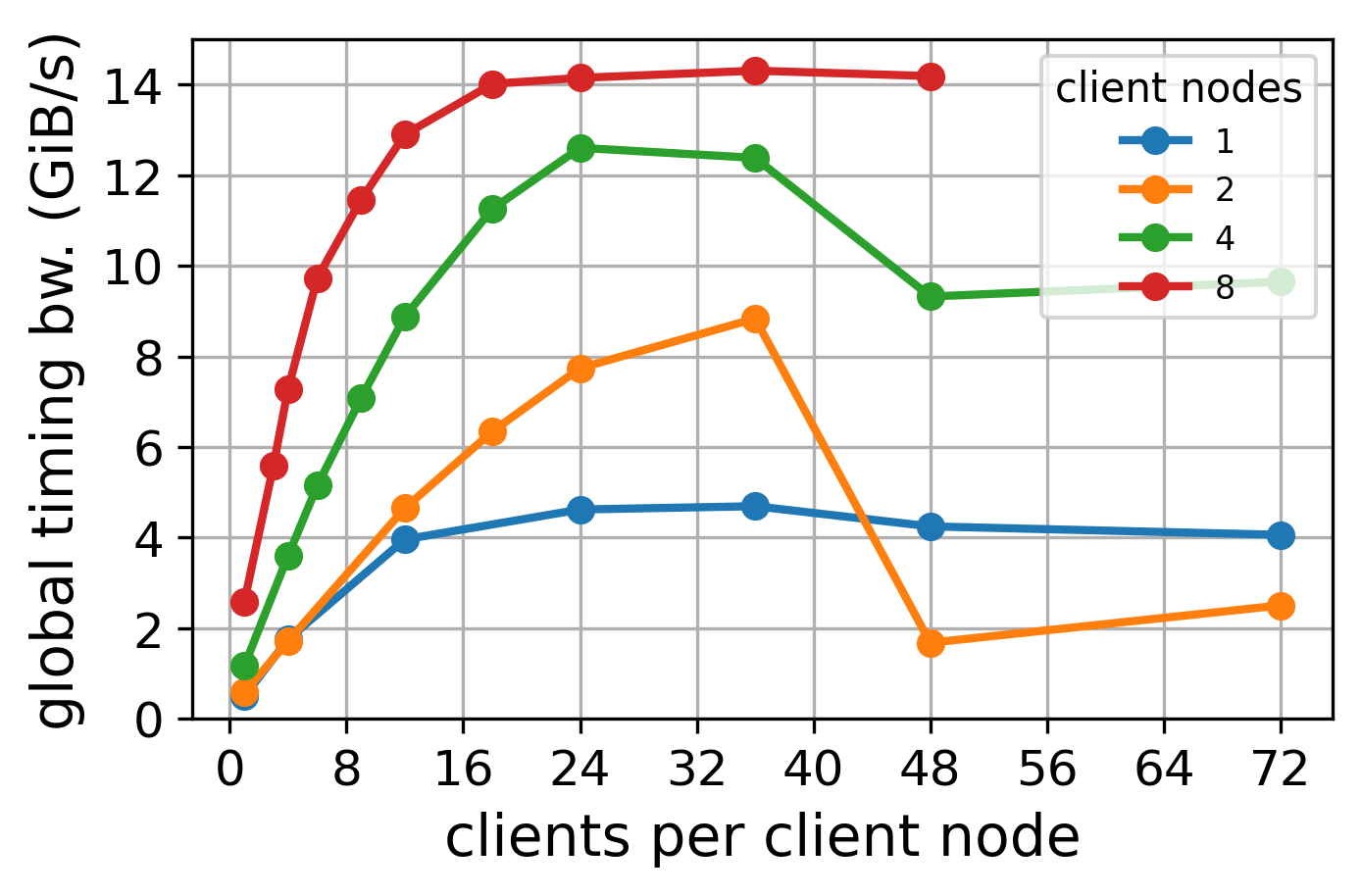}
        \caption{Access pattern A, read}
    \end{subfigure}
    \caption{Global timing write and read bandwidth results for access pattern A with the Field I/O benchmark, with two DAOS server nodes.}
    \label{fig:fieldio_cn_cpcn}

\end{figure}

The results show that, both for write and read, using double the amount of client nodes as server nodes is a good trade-off, reaching high bandwidths without employing the largest amount of resources, and 36 or 48 processes per client node usually result in the best performance. Further variations of the experiment have shown that these results apply as well to pattern B and other server counts, and therefore these configuration values have been used for most of the benchmark runs reported below.

\subsubsection{Scalability of the Field I/O modes}

Access patterns A and B have been run with all modes of the Field I/O benchmark, configured for maximum contention on the indexing Key-Values. DAOS has been deployed on varying numbers of server nodes (1 to 8). Benchmarks have been repeated 10 times, using double the amount of client nodes, 36 and 48 processes per client node, and the same values for object class, object size and repetitions as above. The averaged results for the best configurations are shown in Fig. \ref{fig:fieldio}.

\begin{figure*}[htbp]
    \centering
    \begin{subfigure}[b]{120pt}
        \includegraphics[width=120pt]{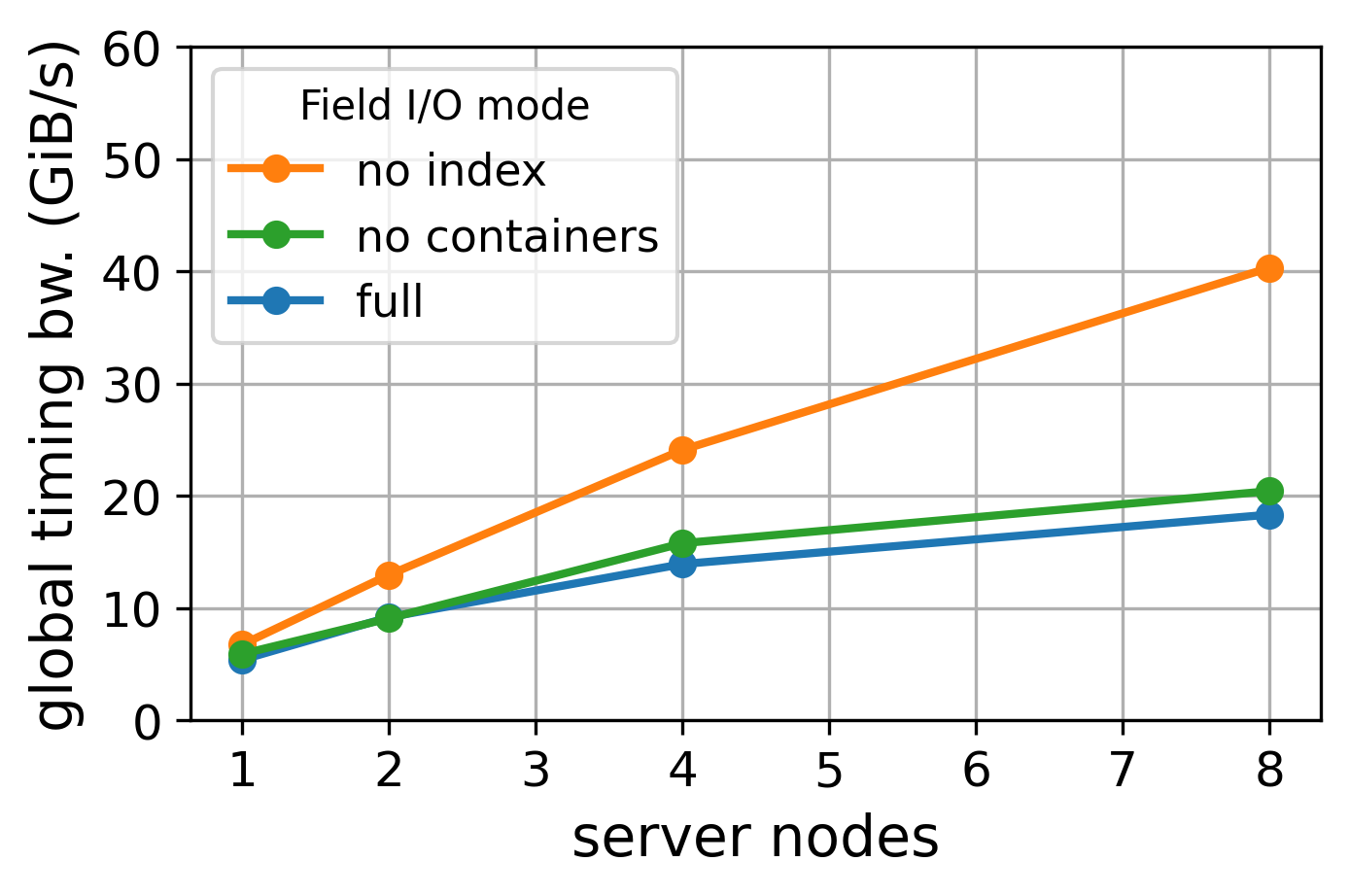}
        \caption{Access pattern A, write}
    \end{subfigure}
    \begin{subfigure}[b]{120pt}
        \includegraphics[width=120pt]{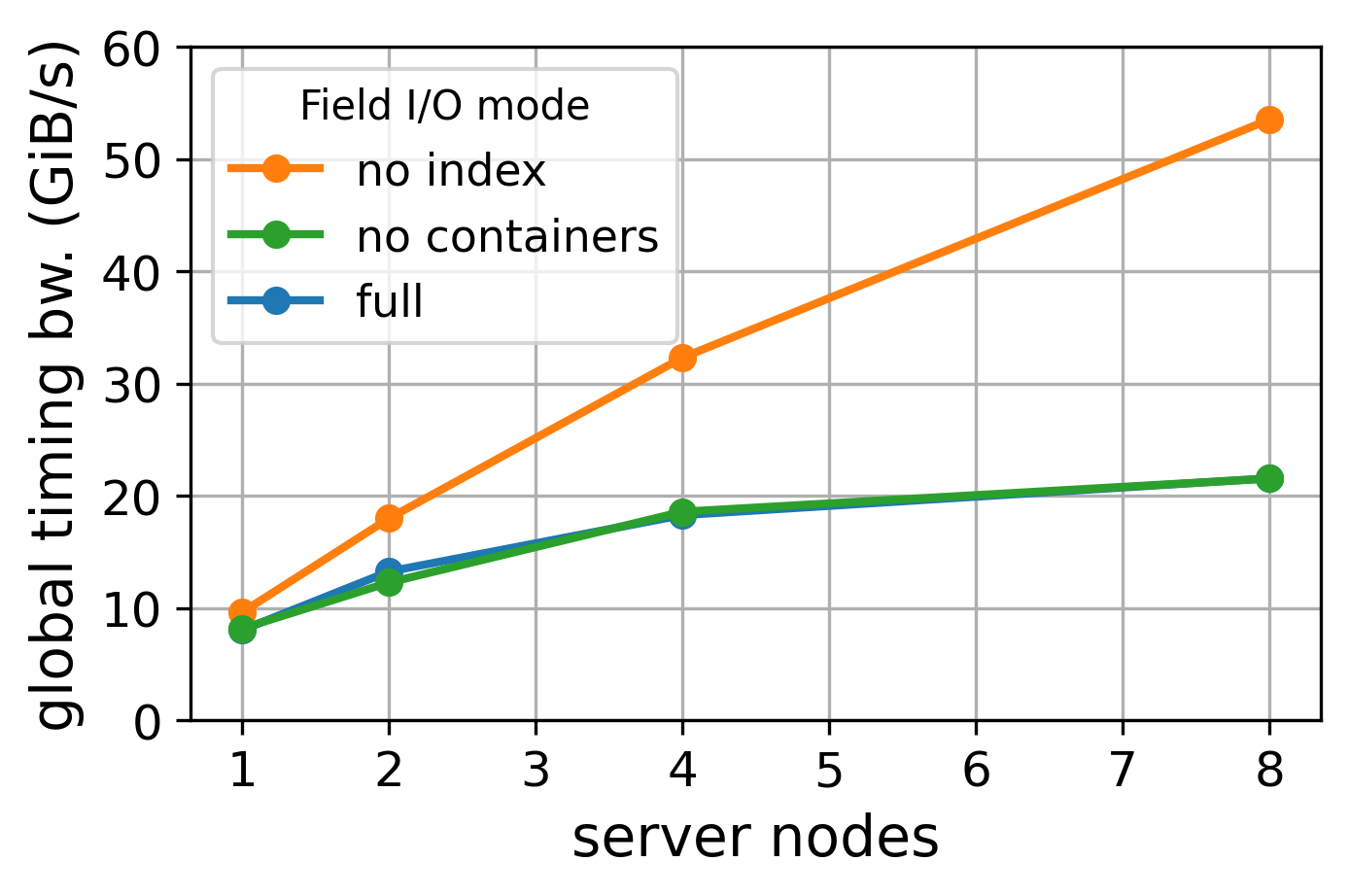}
        \caption{Access pattern A, read}
    \end{subfigure}
    %\vskip\baselineskip
    \begin{subfigure}[b]{120pt}
        \includegraphics[width=120pt]{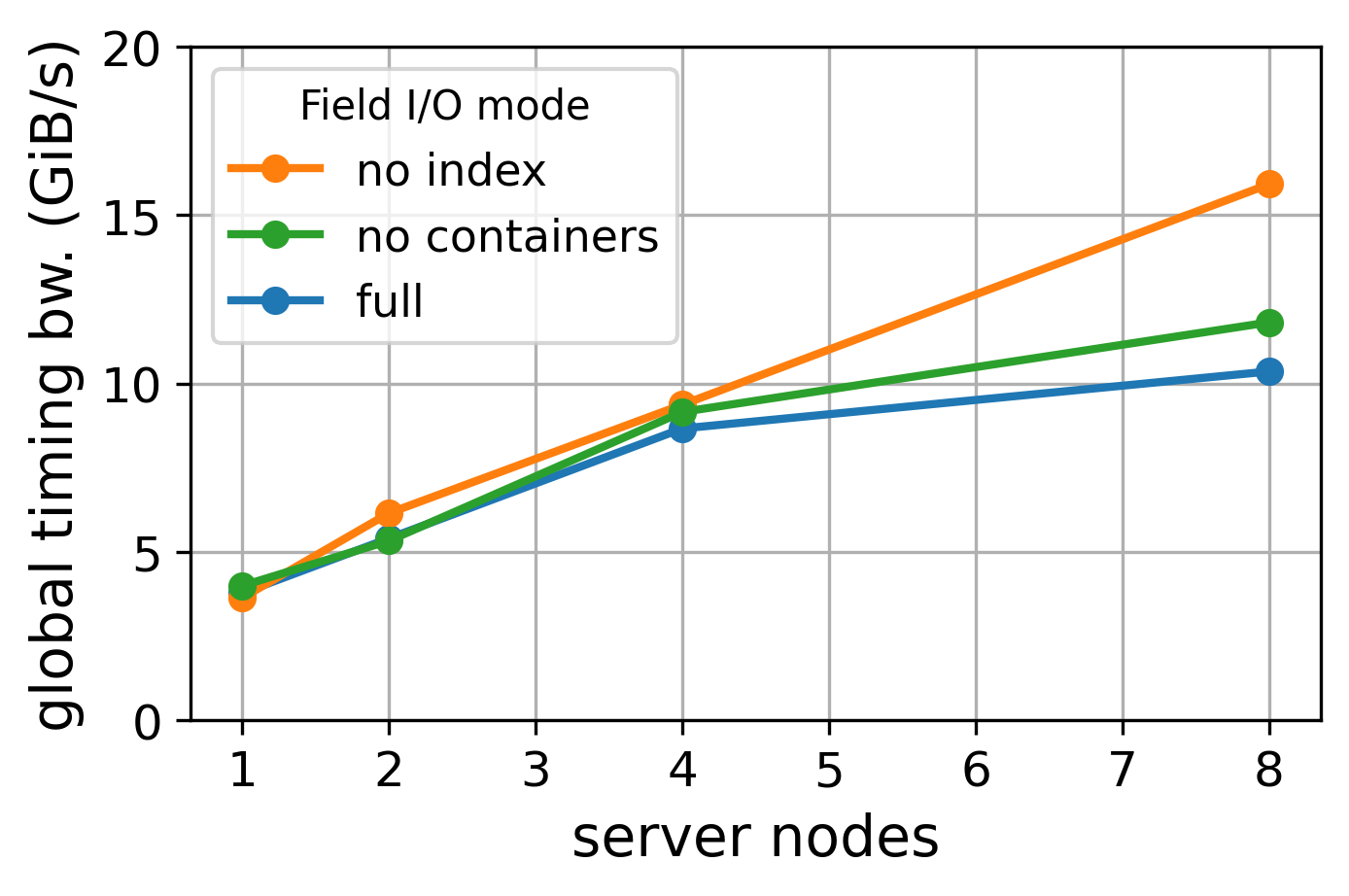}
        \caption{Access pattern B, write}
    \end{subfigure}
    \begin{subfigure}[b]{120pt}
        \includegraphics[width=120pt]{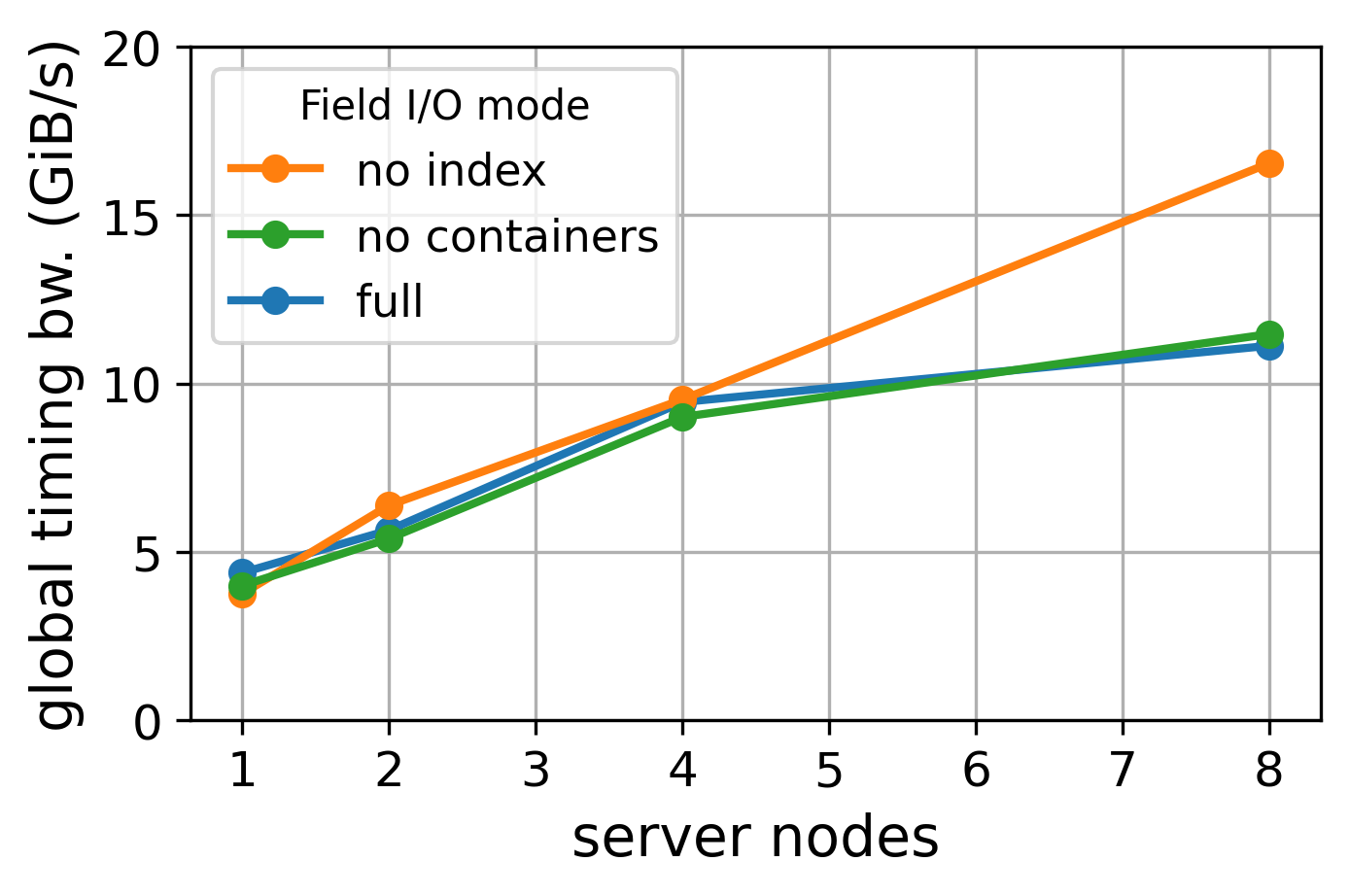}
        \caption{Access pattern B, read}
    \end{subfigure}
    \caption{Global timing write and read bandwidth results for access pattern A (unique writes then unique reads) and B (repeated writes while repeated reads) with the Field I/O benchmark, with high contention on the Key-Value objects.}% The bandwidth scales well, and there is no performance degradation in pattern B.}
    \label{fig:fieldio}

\end{figure*}

We can see from Fig. \ref{fig:fieldio} that the bandwidths obtained in the mode without indexing are of the same order of magnitude and often slightly higher than those observed with IOR. This is likely due to the fact that Field I/O operations are spread over the duration of the benchmark run, rather than being dispatched simultaneously at every iteration as in IOR, allowing the storage to absorb them gradually. Also, the large object and transfer sizes involved in IOR may impact IOR bandwidths, although further experiments in object size are required to confirm this observation. In the full and no containers modes the bandwidths are generally lower than in the mode without indexing and IOR, as expected, due to the increased complexity and number of storage operations. 

The graphs demonstrate that bandwidth scales as the number of server nodes is increased, even with the high degree of contention, with all Field I/O implementations and access patterns, with a slightly better scaling in access pattern B.

Note that in access pattern B the writers and readers run simultaneously, and therefore the write and read bandwidths should be aggregated as an approximate measurement of overall application throughput. It can be observed, if calculating the aggregated bandwidths, that there is no substantial performance degradation in access pattern B compared to pattern A, which is another encouraging result for the DAOS object store with I/O patterns such as the NWP use case.

The simplified mode of the Field I/O functions without indexing (no indexing Key-Values or containers) scales better than the two other implementations, at a rate of approximately 2.5 GiB/s for write and 3 GiB/s for read, per engine, in access pattern A. This scaling rate is similar to that observed with the IOR benchmarks, and demonstrates the capability of DAOS to perform and scale well even when, in contrast to IOR segments, separate I/O operations are issued for the different data parts managed by the client processes. In access pattern B, the increase in aggregated bandwidth for that mode is of approximately 2 GiB/s per engine, which we consider to be strong performance given the read/write contention. 

For the modes with indexing, we observe the use of DAOS containers does not have a substantial impact in Field I/O performance in this scenario. These modes scale up to 4 server nodes at a rate of approximately 1.5 GiB/s of write and read bandwidth per engine for pattern A, and 2 GiB/s of aggregated bandwidth per engine in pattern B, however the scaling rate decreases beyond 4 server nodes down to 0.5 GiB/s of aggregated bandwidth per engine.

The results shown so far have been obtained with Field I/O configured with a single shared forecast indexing Key-Value, inducing very high contention. This is however a very pessimistic scenario, and such a degree of contention is unlikely to occur in operational workloads. The complete test set has been re-run with lower contention, where each client process uses its own forecast index Key-Value, to mimic an optimistic usage scenario. The results, obtained for up to 16 server nodes, are shown in Fig. \ref{fig:fieldio_no_contention}.

The same benchmark configuration as in the runs with high contention have been applied, also using double the amount of client nodes where possible. Due to resource limits, only 20, 14 and 16 client nodes have been used for runs with 12, 14 and 16 server nodes, respectively. These special configurations have been represented with hollow dots in the figure and, whilst the bandwidths they show are probably not as high as they would be if using double as many client nodes, they indicate a lower bound for the scalability curves at high server node counts.

\begin{figure*}[htbp]
    \centering
    \begin{subfigure}[b]{120pt}
        \includegraphics[width=120pt]{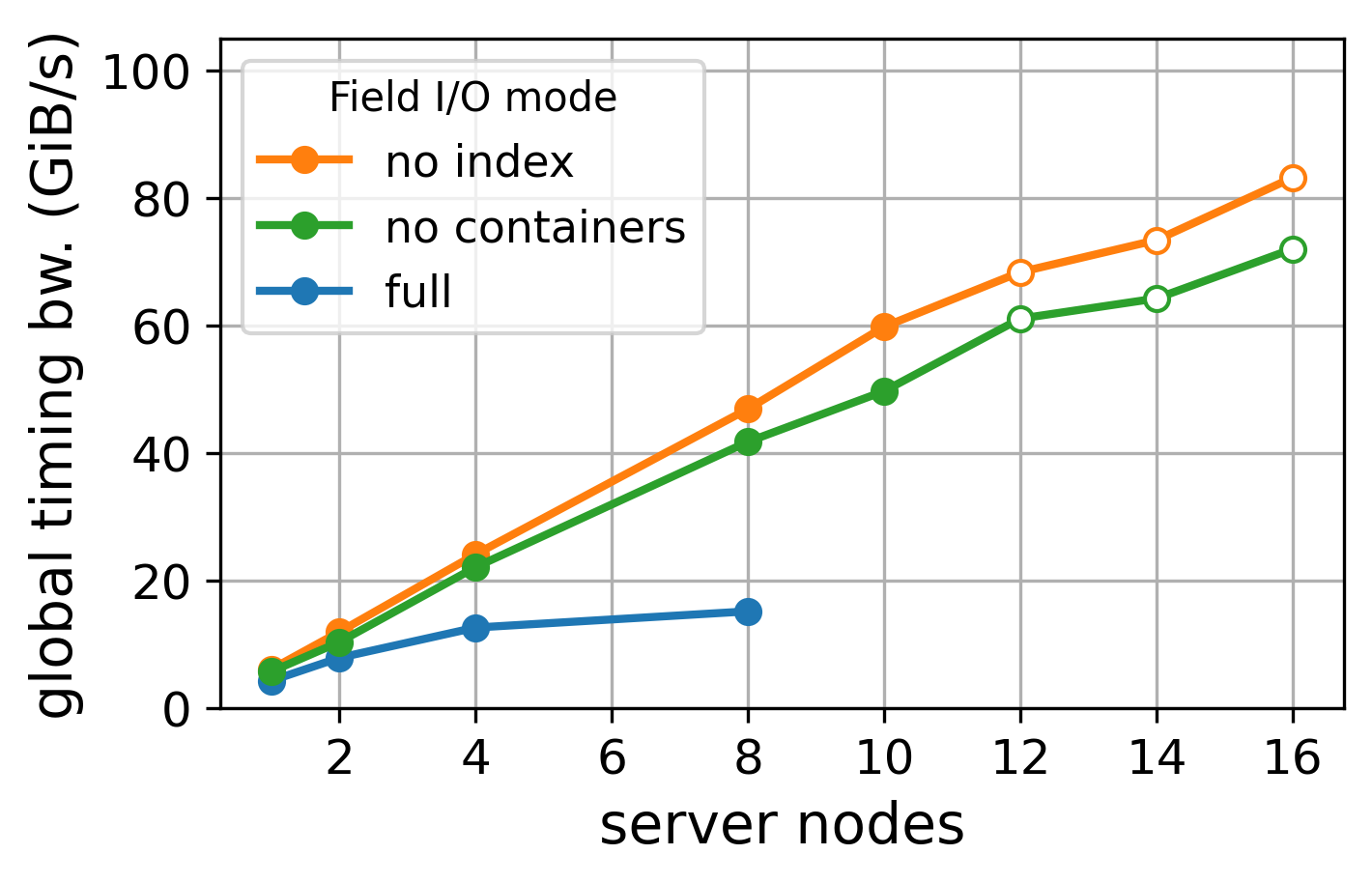}
        \caption{Access pattern A, write}
    \end{subfigure}
    \begin{subfigure}[b]{120pt}
        \includegraphics[width=120pt]{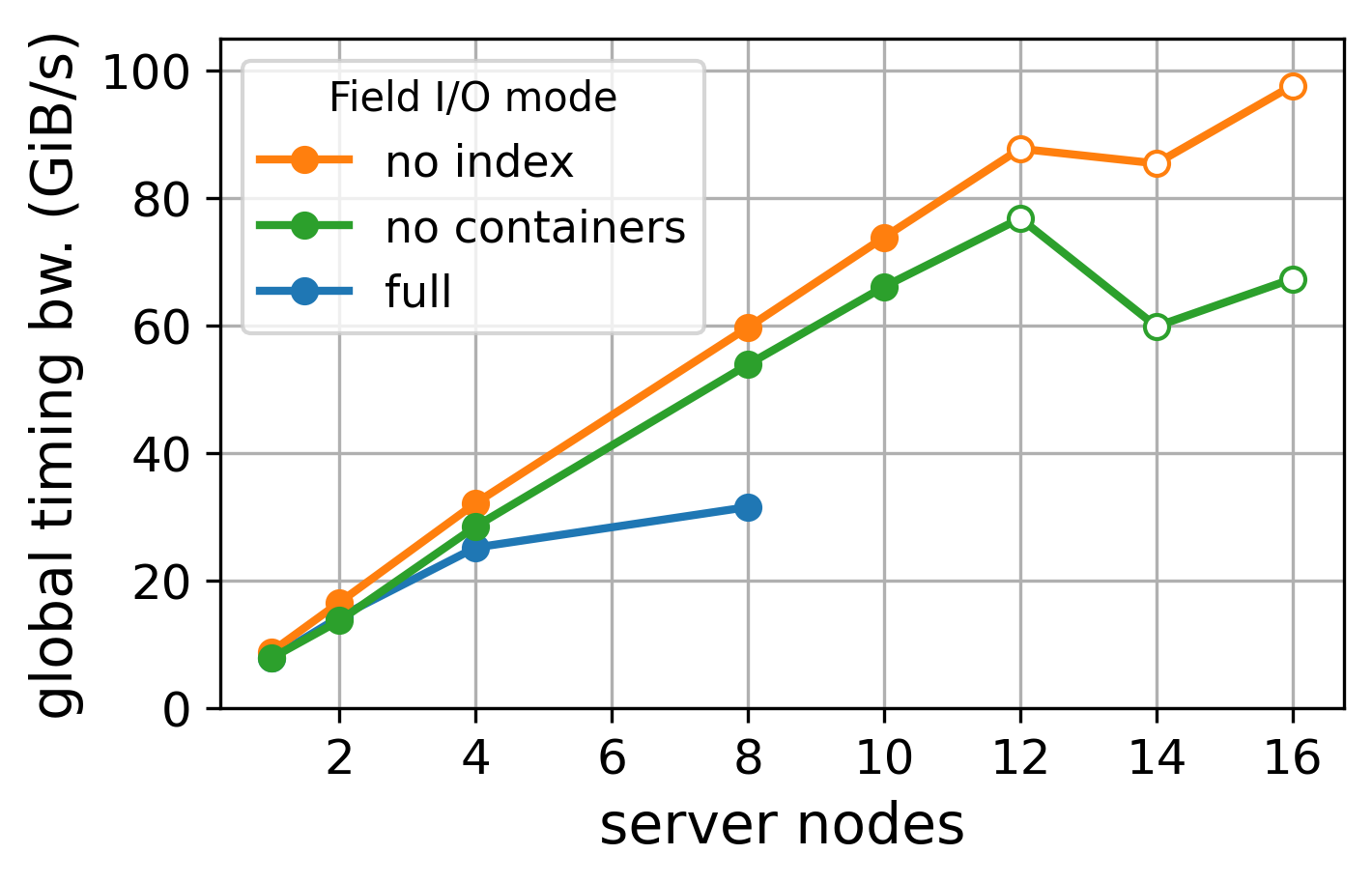}
        \caption{Access pattern A, read}
    \end{subfigure}
    %\vskip\baselineskip
    \begin{subfigure}[b]{120pt}
        \includegraphics[width=120pt]{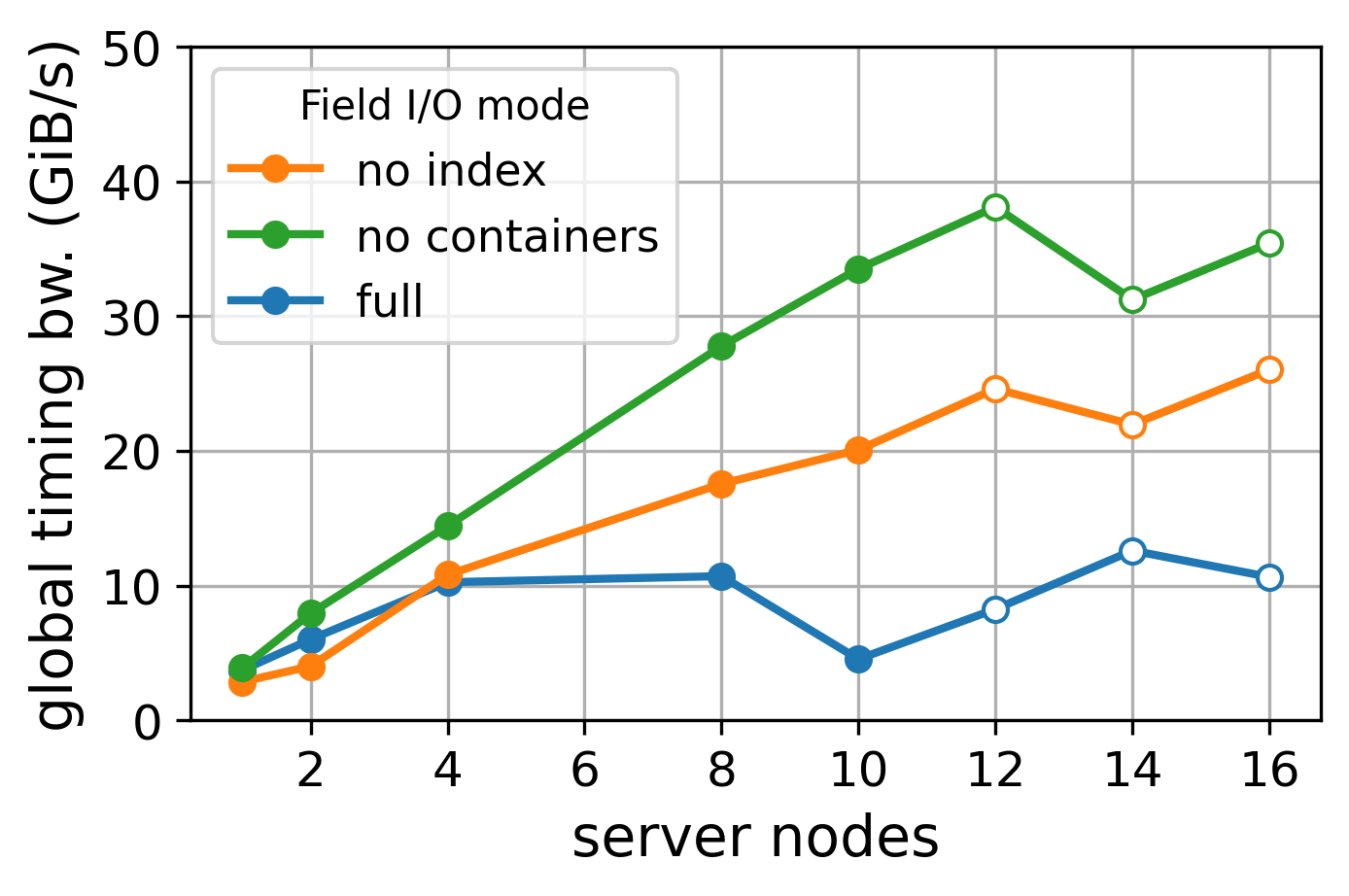}
        \caption{Access pattern B, write}
    \end{subfigure}
    \begin{subfigure}[b]{120pt}
        \includegraphics[width=120pt]{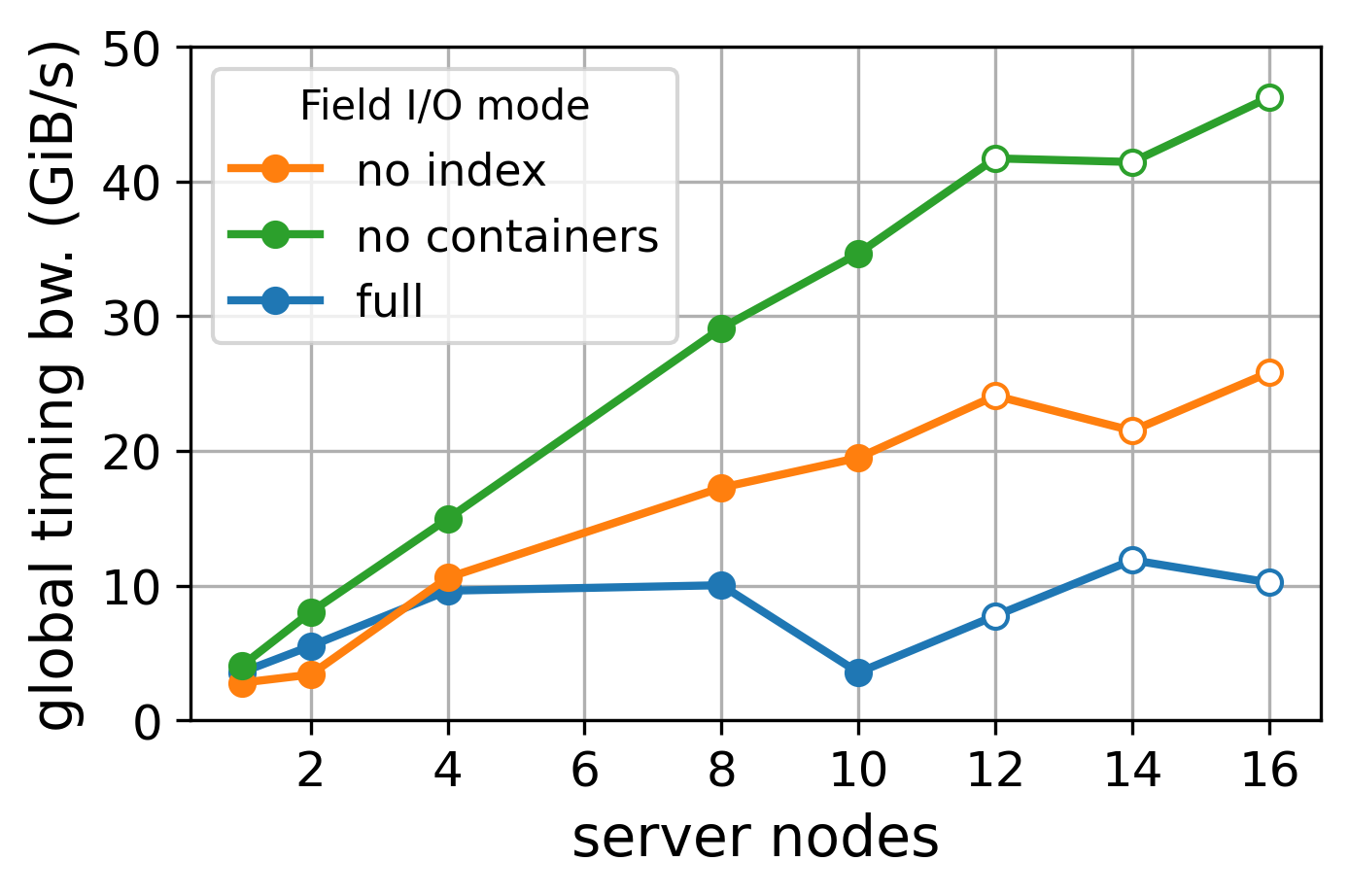}
        \caption{Access pattern B, read}
    \end{subfigure}
    \caption{Global timing write and read bandwidth results with the Field I/O benchmark, with low contention on the Key-Value objects. Hollow dots show configurations where less than twice as many client nodes as server nodes were employed.}% The bandwidth scales very well, particularly if multiple containers are avoided.}
    \label{fig:fieldio_no_contention}
\end{figure*}

%In Figure \ref{fig:fieldio_no_contention}, configurations where the number of client nodes employed was less than twice the number of server nodes, have been represented with hollow dots. For the rest of the configurations, where twice the amount of client nodes or more were employed, are marked with solid dots.

The encouraging results for access pattern A show that the Field I/O implementation without containers scales remarkably well along with the mode without indexing. The full mode performs significantly worse, although it reaches higher bandwidths than in pattern A with high contention.

For access pattern B, the mode without indexing scales at a rate of approximately 2 GiB/s of aggregated bandwidth per additional engine. The full mode does not scale beyond 4 server nodes. The poor performance obtained with this mode suggests that there may be issues with the use or implementation of containers at scale, which needs further exploration.

The mode without containers in pattern B stands out, scaling at a rate of approximately 3.2 GiB/s of aggregated bandwidth per additional engine. With this mode, employing 12 server nodes, a total aggregated application bandwidth of approximately 80 GiB/s is achieved. These encouraging results demonstrate the potential of object storage and DAOS for HPC applications that can exploit such I/O patterns.

For comparison, our reference weather forecasting centre currently uses a Lustre file system, with approximately 300 Lustre Object Storage Target (OST) nodes, each with 10 spinning disks of 2 TiB. It provides a file-per-process IOR bandwidth of up to 165 GiB/s, and a sustained application bandwidth in the order of 50 GiB/s during a typical model and product generation execution, accounting for both the write and read workload. %IOR is described in the Methodology section.
%Further work will be necessary to investigate the cause of the low performance obtained with the Field I/O mode with containers, to address any flaws or optimise the use of DAOS containers.

\subsubsection{I/O distribution results}

I/O offsets have been measured for all Field I/O scalability runs, as described in Asynchronous I/O distribution.

Due to the Field I/O benchmark being configured to synchronise processes at the start and perform a large number of I/O iterations per process, $off_0$ and $po_0$ have been measured to consistently range between 0\% and 2\% of the total parallel I/O wall-clock time, for all access patterns, Field I/O modes and server node counts. 

The measured values for $off_f$ have been found to range between 10\% and 50\% both for write and read phases in runs of access pattern A with different Field I/O modes and at different scales. For access pattern B, $off_f$ ranged between 20\% and 90\%. Values for $po_f$ measured across all runs of access pattern B ranged between 0\% and 20\%.

The large $off_f$ values suggest that I/O operations within a phase were processed in an unbalanced manner. This could be due to the synchronisation mechanism implemented in Field I/O ($off_0$ values are not always 0\%), enabling some processes to initially undertake I/O with low contention from other processes, or due to how DAOS handles I/O requests.

On the other hand, lower $po_f$ values, together with low $off_0$ and $po_0$ values, indicate that the I/O operations in the experiments were distributed as intended according to the desired access patterns.

\subsubsection{Object class and size}

% As previously discussed, there are numerous ways that DAOS can be configured when creating the system or containers within the system. One of the key user mechanisms for configuration is setting the object class associated with a given object. We investigated varying object classes, which can be considered similar to striping of data in a system such as Lustre, or distributing objects to different DAOS engines, depending on the object sizes. For this benchmarking we also varied the object size, to evaluate the performance impact of I/O size on DAOS.
We also investigated the effect of object granularity and different striping configurations for the Field I/O objects. Fig. \ref{fig:object_class_bandwidth} shows bandwidth results from runs of access pattern A with Field I/O in its mode without containers, with object sizes 1, 5, 10, 20 and 50 MiB, and object configurations ranging from no striping (\verb!OC_S1!) to striping all objects across all targets (\verb!OC_SX!). Field I/O has been configured to run with no contention in the indexing Key-Values.

All tests have been run with a fixed configuration of 2 server nodes and 4 client nodes, with 36 processes per client node and 100 I/O operations per client process. Benchmarks are repeated 10 times for a range of client process counts. The results for the 10 repetitions have been averaged, with the average bandwidth for the highest performing number of client processes per client node shown for each combination of object class and I/O size tested.

\begin{figure}[htbp]
    \centering
    \begin{subfigure}[b]{200pt}
        \centering
        \includegraphics[width=200pt]{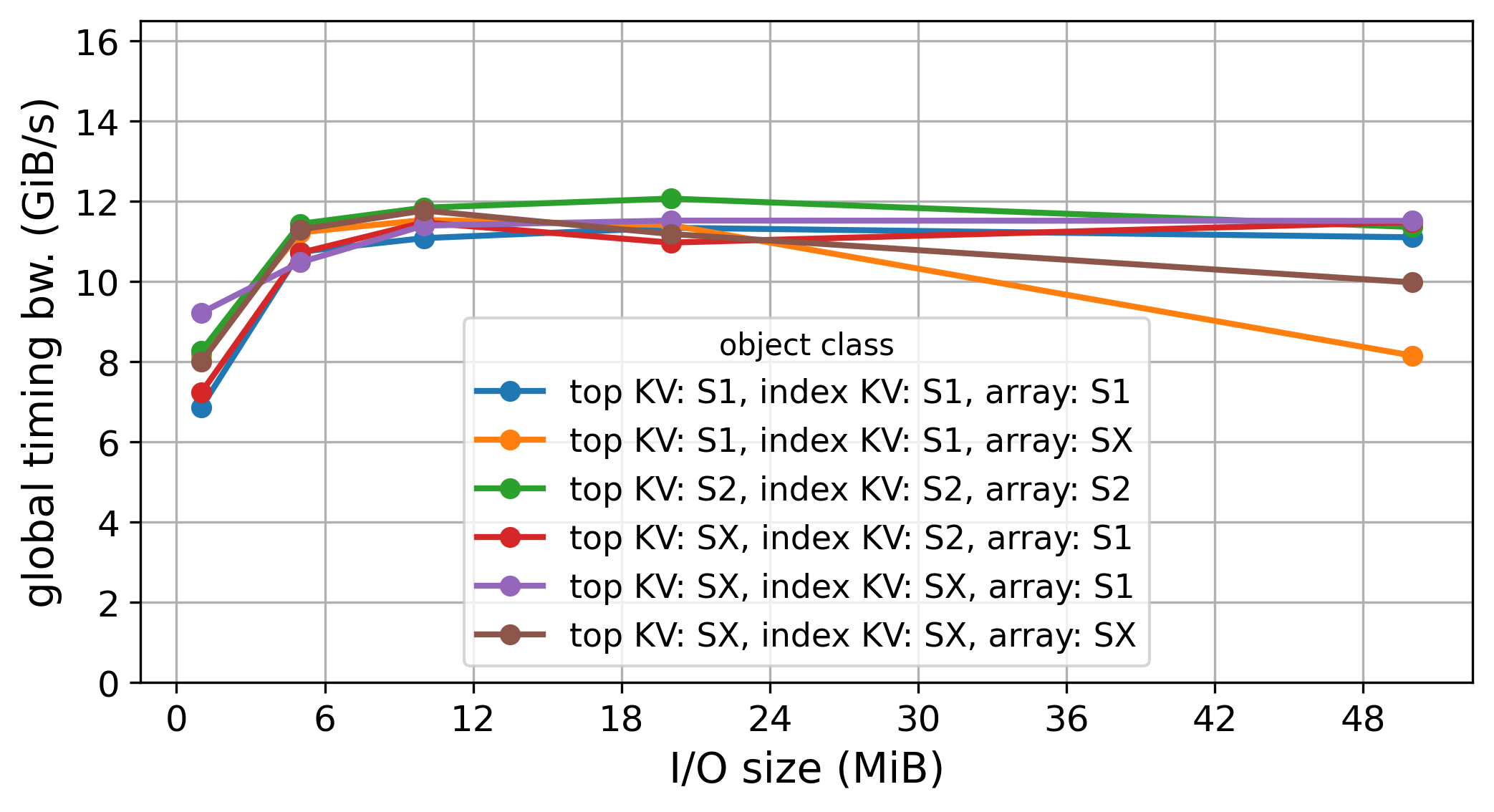}
        \caption{Write}
    \end{subfigure}
    %\vskip\baselineskip
    \begin{subfigure}[b]{200pt}
        \centering
        \includegraphics[width=200pt]{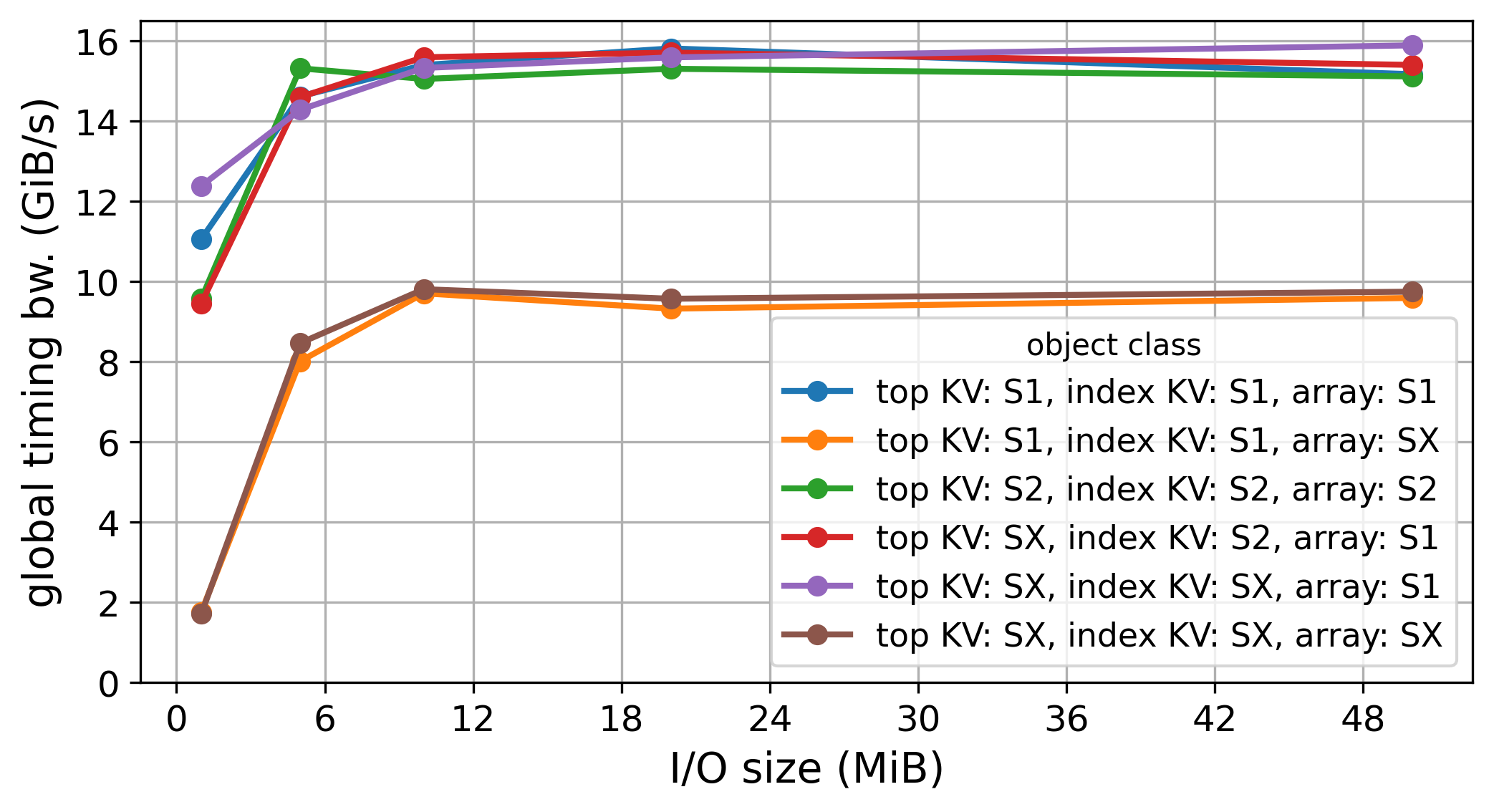}
        \caption{Read}
    \end{subfigure}
    \caption{Global timing write and read bandwidth results for access pattern A, with Field I/O in ``no containers" mode, using 2 server and 4 client nodes.}% Increasing field Array size above 1 MiB has a positive impact. Striping all objects across all targets is beneficial for write, while striping across two targets is beneficial for read.}
    \label{fig:object_class_bandwidth}
\end{figure}

It can be observed that, whilst all object class configurations result in similar performance, increasing Array object sizes from 1 to 5 MiB or 10 MiB has a substantial benefit in I/O performance for both write and read. This is a positive result for the NWP use case, which indicates that as higher-resolution data is used in the future, scaling will improve rather than deteriorate.

For configurations with no or some striping (\verb!OC_S1! and \verb!OC_S2!) on the Array objects, the bandwidth plateaus or drops slightly beyond 10 MiB. This suggests that the IOR benchmarking, where we used 100 MiB objects, may not provide optimal results, although further benchmarking is required to confirm that. In configurations with maximum striping (\verb!OC_SX!) on the Array objects, the read performance is consistently lower than with other configurations for all object sizes, and the write performance is similar at small object sizes but declines substantially beyond 10 MiB. This is not surprising as striping is usually beneficial when parallel processes operate with shared objects, but DAOS objects are not shared in the Field I/O modes with indexing and without contention.

We note that all other Field I/O benchmark runs in this paper have been configured to use non-striped 1 MiB Arrays and maximum striping for the Key-Values, which Fig. \ref{fig:object_class_bandwidth} demonstrates is one of the best-performing configurations.

The bandwidths for access pattern B when varying object class and size, not included here, show similar behaviour, with the difference being that write and read bandwidths range from 5 to 9 GiB/s. Re-running the benchmark with larger amounts of server and clients nodes also showed similar behaviour, with no decline in write performance beyond 10 MiB, and a more notable decline in read performance beyond 10 MiB.

\section{Conclusions}

DAOS has been demonstrated to provide object storage functionality meeting the demands of the weather field store currently in use for time-critical operations at the ECMWF, our exemplar weather forecasting centre. For this study, a weather field I/O benchmark has been developed and employed to evaluate if DAOS provides the required levels of consistency, functionality, and stability.

The performance of DAOS in conjunction with NVM has been tested in the NEXTGenIO research HPC system using the well-known IOR benchmark and the developed field I/O benchmark in a variety of modes and access patterns. The results demonstrate the capability of DAOS to scale linearly in throughput in applications for storage and indexing of weather fields or other relatively small data objects.

The aggregated bandwidth achieved in such applications increases at a rate of up to 3.2 GiB/s per additional socket with SCM in the HPC system. Using up to 12 server nodes and 20 client nodes, the aggregated bandwidth reaches up to 80 GiB/s, using a sub-optimal object size. For comparison, we used a 300-node Lustre file system at our reference weather centre, which provides an aggregated bandwidth on the order of 50 GiB/s during a typical operational run.

These encouraging results, although qualified by a limited number of server and client nodes, indicate that a small DAOS system with non-volatile storage, on the order of a few tens of nodes, could perform as well as the HPC storage currently used for operations at the ECMWF, and suggest DAOS has the potential to support the next generation of weather models that will generate significantly larger amounts of data and require much larger I/O bandwidth and more intensive metadata operations. 
%, which require much larger I/O bandwidth and metadata operations to support efficient data creation and usage.

%It has also been demonstrated that performance improves as the object size increases beyond 1 MiB, so as we move to higher resolution simulations in the future our scaling will improve rather than deteriorate. Nevertheless, further exploration is required to assess the impact of object granularity, particularly with substantially larger object sizes.

The demonstration of strong I/O performance using a standard IOR configuration benchmarking DAOS also demonstrates that applications not adopting object store approaches or semantics are still likely to obtain good performance on a DAOS storage system, highlighting the applicability of DAOS for general HPC storage in future systems.

We note that our configuration of deploying DAOS entirely on SCM, rather than a combination of SCM and NVMe devices, may have had a positive impact on the performance DAOS achieved, but speculate that some of this benefit has been reduced by the network fabric restrictions (TCP rather than a higher-performance fabric provider) that the benchmarking has been undertaken with. %We also note that any I/O benchmarking will be affected by the underlying hardware being used for the server nodes, the client nodes, and the network between them, so this study is not unique in relying on the specific hardware configuration chosen for benchmarking.

%Whilst this paper has broadly shown that DAOS is ready for production workloads, we encountered some issues that impacted some tests during the benchmarking. For instance, our benchmarks with Field I/O in full mode, access pattern A with low contention failed using more than 8 server nodes. This benchmarking exercise has been beneficial in discovering bugs or issues within DAOS and helping the DAOS developers move to a fully hardened, production-ready storage system.

The introduced Field I/O framework opens a new area of benchmarking of high-performance object stores to assess their potential to enable I/O for next-generation NWP models. As part of this framework, we have formalised a network bandwidth measurement, global timing bandwidth, that we think more accurately represents the bandwidth achievable by mixed workloads on a shared I/O system. %This has enabled us to more accurately assess the bandwidth our target applications could actually experience when deployed on a production system, and as such helps us move from I/O benchmarking that really only captures ``best possible" performance, to benchmarking that represents achievable or realistic performance.
We consider that such benchmarking metrics will be important for ensuring user experience matches vendor deployments and procurement objectives for future I/O systems.

This preliminary assessment has showcased an example of what performance characteristics and design and configuration issues may be encountered when migrating from a high-performance distributed file system with traditional storage hardware to a high-performance object store with NVM storage, as well as the implementation of a domain-specific object store. It has shown the implementation of domain-specific object stores can result in performance benefits in certain I/O scenarios, and has helped investigate the specific requirements and the design of appropriate data layouts to use DAOS for such object stores. %The results are sufficiently promising to justify the much larger future work of integrating a DAOS back-end into a production I/O library and testing at a larger scale with real NWP simulation workloads.

%There are some places where the analysis that has been documented
%The analysis in this paper can be further extended or refined, including investigating DAOS performance with larger numbers of server nodes, client nodes, and client processes, and implementing and benchmarking direct DAOS interfaces for production NWP applications. These would help to build confidence in this initial assessment that DAOS has the potential to provide significantly enhanced functionality and performance for future applications, particularly in the field of weather forecasting.

% to explain why some of the performance measurements obtained with the field I/O benchmark are higher than the IOR counterpart.

% Repeat testing with sleeps in field IO then substract!!! This will allow to see how daos performs when clients are not constantly sending IO

% Repeat IOR testing with "transfers" mode -> this will allow to see if there's a substantial change when transfrers are not huge, even though objects are still huge

% Try more runs where the distribution across more KVs is enabled

%Work on adoption of DAOS will be continued by implementing an FDB5 back-end, testing workloads more similar to those in the centre's operational workflows, and possibly employing more server nodes to ensure DAOS scales in performance in setups one or two orders of magnitude larger.

\pagebreak

\section*{Acknowledgment}

The work presented in this paper was carried out with funding by the European Union under the Destination Earth initiative (cost center DE3100, code 3320) and relates to tasks entrusted by the European Union to the European Centre for Medium-Range Weather Forecasts. Views and opinions expressed are those of the author(s) only and do not necessarily reflect those of the European Union or the European Commission. Neither the European Union nor the European Commission can be held responsible for them.

The NEXTGenIO system was funded by the European Union's Horizon 2020 Research and Innovation program under Grant Agreement no. 671951, and supported by EPCC, The University of Edinburgh.

%\vspace{40px}

%\vspace{20px}

\vspace{-6 mm}
\begin{IEEEbiography}[{\includegraphics[width=1in,height=1.25in,clip,keepaspectratio]{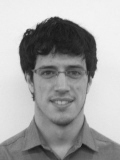}}]{Nicolau Manubens}
obtained a licentiate in informatics from the Universitat Autònoma de Barcelona in 2014, and has since worked as a software engineer at various weather and climate forecasting and research institutions. He is currently employed at the European Centre for Medium-Range Weather Forecasts and is a Ph.D. candidate at EPCC, The University of Edinburgh, doing research work on HPC storage for km-scale Numerical Weather Prediction in the context of the Destination Earth initiative.
\end{IEEEbiography}

\vspace{-6 mm}
\begin{IEEEbiography}[{\includegraphics[width=1in,height=1.25in,clip,keepaspectratio]{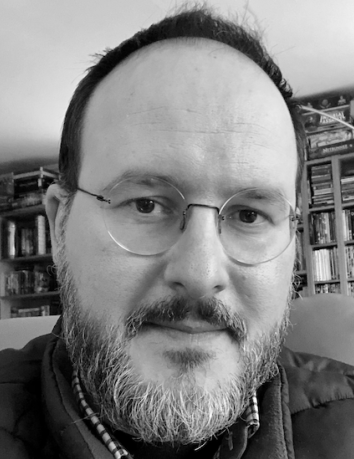}}]{Tiago Quintino}
is Head of Software Development section at the ECMWF, where they develop high-throughput specialist software supporting ECMWF's operational activity. %meteorological forecast model, systems for acquisition of incoming observations, management of direct model output, perpetual archival of weather observations and forecast data, and post-processing, generation, visualisation and dissemination of meteorological products. They also develop cloud meteorological and climate data provisioning services (Data-as-a-Service) in support of ECMWF's cloud activities, as web services for accessing and visualising such data.
Dr. Quintino's career spans 20 years researching numerical algorithms and developing high-performance scientific software in the areas of Aerospace and Numerical Weather Prediction. Lately, his research focuses on scalable data handling algorithms for generation of meteorological forecast products, optimising the workloads and I/O of massive data sets towards Exascale computing.
\end{IEEEbiography}

\vspace{-6 mm}
\begin{IEEEbiography}[{\includegraphics[width=1in,height=1.25in,clip,keepaspectratio]{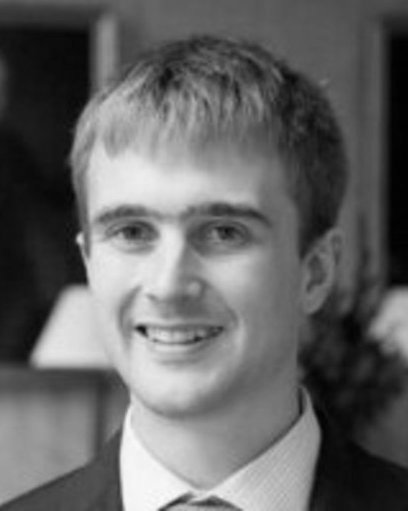}}]{Simon Smart}
obtained a Ph.D. in theoretical chemistry in 2013 from the University of Cambridge, and then worked at the Max Planck Institute for Solid State Chemistry writing high-performance computational chemistry software. He has worked at the European Centre for Medium-Range Weather Forecasts since 2015, focusing on scalability and the challenges of data handling on the pathway to Exascale. He now leads the Data Management Services team, with responsibility for all of the software infrastructure underpinning data handling at ECMWF.
\end{IEEEbiography}

\vspace{-6 mm}
\begin{IEEEbiography}[{\includegraphics[width=1in,height=1.25in,clip,keepaspectratio]{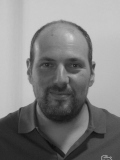}}]{Emanuele Danovaro}
received a M.Sc. (in 2000) and a Ph.D. (in 2004) in computer science from the University of Genova. During his post-doc, he developed algorithm and compact data structures for multi-dimensional data. In 2009, he joined the Italian National Research Council as a senior researcher, working on algorithm parallelisation and leading the development of a workflow manager for meteorological applications. In 2019 he joined ECMWF and is currently responsible for the development of the high-performance data management subsystem.
\end{IEEEbiography}

\vspace{-6 mm}
\begin{IEEEbiography}[{\includegraphics[width=1in,height=1.25in,clip,keepaspectratio]{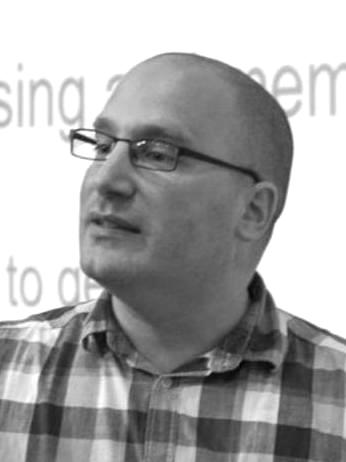}}]{Adrian Jackson}
is a Senior Research Fellow at EPCC, The University of Edinburgh. He has a history of research into high-performance and parallel computing, with a focus on parallel algorithms, domain decomposition, parallel I/O, and memory technologies. He collaborates with a wide range of domain scientists, from plasma physicists and radio astronomers, to CFD experts and biologists.
\end{IEEEbiography}

\end{document}